\documentclass[reprint, superscriptaddress, aps, showkeys, pra]{revtex4-2}

\usepackage[colorlinks=true, citecolor=blue, urlcolor=blue, linkcolor=magenta]{hyperref}
\usepackage{braket}
\usepackage{amsmath,amssymb,amsthm}
\usepackage{easybmat}
\usepackage[colorlinks=true,citecolor=blue,urlcolor=blue, linkcolor=magenta]{hyperref}
\usepackage[pdftex]{graphicx}
\usepackage{times,txfonts}
\usepackage{braket}
\usepackage{color}
\usepackage{natbib}
\usepackage{appendix}

\begin{document}
    \title{Quantum Chaos in Non-Markovian Open Quantum Systems: Interferometric OTOC, Loschmidt Echo and Commutator Operator Norm}
\author{Baibhab Bose\textsuperscript{}}
\email{baibhab.1@iitj.ac.in}
\author{Devvrat Tiwari\textsuperscript{}}
\email{devvrat.1@iitj.ac.in}
\author{Subhashish Banerjee\textsuperscript{}}
\email{subhashish@iitj.ac.in}
\affiliation{Indian Institute of Technology Jodhpur-342030, India\textsuperscript{}}

\date{\today}

\begin{abstract}
    Out-of-time order correlators (OTOCs) are crucial tools for studying quantum chaos as they show distinct scrambling behavior for chaotic Hamiltonians. We calculate OTOC and analyze the quantum information scrambling in atom-field and spin-spin interaction models, which are open-system models and exhibit non-Markovian behavior. We also examine the Loschmidt echo for these models and comment on their chaotic nature. The commutator growth of two local operators, which is upper bounded by the Lieb-Robinson bound, is studied for these models, and the patterns of scrambling are investigated.
\end{abstract}

\maketitle
\textbf{Information scrambling is a prominent approach for probing chaos in quantum systems, where the classical notion of sensitivity to the initial condition as the definition of chaos is reinterpreted. In particular, we study quantum chaos using interferometric OTOC, Loschmidt echo, and the operator norm of the appropriate commutator. Specifically, the nature of information scrambling in an Ising spin chain system with nearest-neighbor interaction illustrates many key features of a chaotic quantum system. A local operation on one of the spin sites is considered as the perturbation, and the extent of its residue in time is observed for signatures of chaos in the system. Among all the measures through which these key features of quantum chaos are expressed, the out-of-time order correlator (OTOC) has attracted considerable interest in recent times. OTOC is a four-point correlation function of local operators ordered in a non-linear fashion in time. The operator norm of the commutator of two local operators is another mode of observing information scrambling. Its calculation is motivated by the fact that this quantity is bounded by the Lieb-Robinson bound. In a quantum many-body system with local interactions, the Lieb-Robinson bound sets a limit on how quickly information or correlations can spread through the system. This gives rise to an effective light cone, which imposes a notion of causality in non-relativistic quantum systems, much like how the speed of light defines the light cone structure in Minkowski spacetime. In the context of quantum chaos, the Lieb-Robinson bound provides a natural upper bound on the rate at which information can be scrambled. Another quantum reinterpretation of the classical notion of sensitivity to the initial condition can be obtained by the Loschmidt echo. The Loschmidt echo calculates the overlap of two final states evolved by slightly different Hamiltonians from the same initial state. The effect of a small change in the Hamiltonian on this overlap can be analyzed to shed light on the underlying dynamics. Here, our aim is to understand quantum chaos from an open quantum system perspective. To this effect, we consider the light-matter interaction model (Tavis-Cummings model) and the Ising spin chain in a tilted magnetic field. We calculate the above-mentioned quantifiers of quantum chaos for these open system models and comment on the chaotic nature of the system and the non-Markovian nature of the evolution. Non-Markovianity refers to memory effects where the system’s future evolution depends on its past, due to strong or structured coupling with the environment. One of the significant benchmarks of this is the information backflow from the environment to the system, influencing entanglement dynamics and the nature of information scrambling.}
\section{Introduction} 
In the analysis of the behavior of chaotic quantum mechanical systems, different measures of scrambling of quantum information are used extensively~\cite{García-Mata:2023, Swingle_2024, mehta2004random, haake2001quantum, Toda_1988, Pandey_1981, pandey2019quantum, Zurek_1994, gutzwiller1991chaos}. The stark contrast of the behavior of these quantities in chaotic and non-chaotic regimes indicates that chaotic Hamiltonians are fast scramblers of information~\cite{Swingle2018}. Some of the well-known witnesses of quantum chaos are OTOC (out-of-time ordered correlators)~\cite{García-Mata:2023, Swingle_2024, Hashimoto2017, Shenker2014, Standford_2015, Maldacena2016, Ueda_otoc, devvrat_OTOC}, Bipartite OTOC~\cite{Zanardiopenbipotoc1, zanardibipotoc2, Bose_2024}, the commutator growth~\cite{García-Mata:2023}, among others. These measures involve local operators and thus facilitate the probing of the flow of quantum information.
In recent years, works on the black hole information paradox ~\cite{Maldacena2016, Hayden_2007}, and the problem of
quantum thermalization ~\cite{Srednicki_1994, Haake2, Hirsch_2022} have rekindled interest in this field. Random matrix theory has been a traditional approach to quantum chaos ~\cite{mehta2004random, wigner, pandey2019quantum}. 

The OTOC has found many applications in quantum information scrambling and, in turn, quantum chaos~\cite{Hashimoto2017, Swingle_2024, Maldacena2016}. It is a four-point correlation function made of local operators whose order of application is not linear in time. This quirky structure has been seen to be essential in probing the scrambling of quantum information. The OTOC was brought to light in the context of superconductivity ~\cite{Larkin1969QuasiclassicalMI} and was made popular in ~\cite{Shenker2014}. The analysis of the Lyapunov spectrum of the commutator growth in quantum systems with a classical limit has been another effective approach towards quantum chaos~\cite{Hirscheatomfield, Hirsch_2016, Tian2022}.  OTOCs have been extensively used in the study of a number of systems, such as in quantum field theories ~\cite{Standford_2015}, random unitary models~\cite{David_2018}, spin chains ~\cite{Zhang_2019, Fortes_2020, bose2025_LMG_paper}, quantum phase transition~\cite{otoc_floquet}, and quantum optical models ~\cite{Hirscheatomfield, devvrat_OTOC, Mahaveer}, among others ~\cite{García-Mata:2023, Swingle_2024, Swingle2018}. Various forms of the OTOC, such as the regularized and the physical OTOC, were also discussed in light of the fluctuation-dissipation theorem, generalized to the case of the OTOC ~\cite{Ueda_otoc}. To overcome the operator dependence of OTOC, Haar averaging over the unitary ensemble was done to obtain bipartite OTOC~\cite{zanardibipotoc2}. This was studied in the context of open quantum systems in~\cite{Zanardiopenbipotoc1, Bose_2024}.

An interferometric scheme to calculate the OTOC, which we call the $\mathcal{F}$-OTOC, was provided in~\cite{SwingleInfoScram}. It was generalized for open systems, in particular, for the Gorini-Kossakowski-Sudarshan-Lindblad (GKSL) evolution~\cite{Lindblad1976, GKLSpaper}, in \cite{Zhang_2019}. The present work is devoted to the study of $\mathcal{F}$-OTOC in the context of open quantum systems.  

Apart from the approach of information scrambling, another state-dependent quantity that diagnoses whether a Hamiltonian is chaotic or not is the Loschmidt echo~\cite{zanopen39, LoshM_Goussev2012LoschmidtE}. Here, `echo' signifies the overlap of an initial and final state obtained by evolving the initial state forward and then backward equally in time with a mildly perturbed Hamiltonian in the backward evolution. The decay of the Loschmidt echo exhibits a large deviation from the original state when the Hamiltonian is chaotic.
Further, for a system with local interactions, particularly for spin chain models, the information spread is bounded by the Lieb-Robinson bound \cite{Lieb1972, liebRob_Hastings:2010vzr, LiebRobSwingle_PhysRevLett.117.091602, nachtergaele2011adosomethingliebrobinsonbounds, nachtergaele2010liebrobinsonboundsquantummanybody, Naaijkens2017, Hastings_LR_bound}. This provides an upper limit of the operator norm of the commutator between two local operators at different times. This upper bound is an exponential, which involves a constant called the Lieb-Robinson velocity. Lieb-Robinson velocity is thought of as the finite speed at which quantum information travels through the quantum system.

A number of recent works have focused on the scrambling of quantum information in open quantum systems~\cite{Zhang_2019,devvrat_OTOC,zanopen21, zanopen31, Bose_2024, infor_scram_non_Markov, Deffner_info_scram_otoc}. The field of open quantum systems analyzes the behavior of quantum systems, taking into account the influence of their surroundings~\cite {Weiss2011, Breuer2007, Banerjee2018}. In recent years, this subject has progressed significantly~\cite{Omkar2016, Javid_2018, Vacchini_2011, Tiwari_2023, tiwari2024strong}. A way to approach the understanding of the dynamics of open quantum systems is the GKSL formalism~\cite{GKLSpaper, Lindblad1976}. This makes use of the Born-Markov and rotating wave approximations, generating a Markovian evolution. With progress in theory and technology, the effects of non-Markovian evolution are being taken into account~\cite{Hall_2014, Rivas_2014, RevModPhys.88.021002, CHRUSCINSKI20221, banerjeepetrucione, vega_alonso, Utagi2020, kading2025}.  

The aim of this work is to study the information scrambling using $\mathcal{F}$-OTOC, Loschmidt echo, and operator growth in various open system models, particularly highlighting the impact of non-Markovian evolution. The models studied are the Tavis-Cummings (TC) model~\cite{Larson_Dicke}, which is an atom-field interaction model, and the tilted field Ising model (TFIM)~\cite{TiltedMfieldofarul}. They have been shown to exhibit quantum chaotic properties in particular parameter regimes~\cite{Hirscheatomfield, devvrat_OTOC, Bose_2024, TiltedMfieldofarul}.

The plan of the paper is as follows. In Sec. \ref{f_otoc}, the interferometric scheme of OTOC is described for open systems. The TC and TFI models are illustrated in Sec.~\ref{nmgad}, and $\mathcal{F}$-OTOC is calculated and plotted. In Sec. \ref{loshcmidt echo}, the Loschmidt echo is introduced and calculated for these models. Motivated by the Lieb-Robinson bound, the operator norm of the commutator growth is calculated in Sec.~\ref{operator growth}. The work concludes in Sec.~\ref{conclusion}.

\section{OTOC in (non)-Markovian open quantum systems}\label{f_otoc}
In a quantum many-body system of spins, a local Hamiltonian having, for example, nearest-neighbor interactions makes any initial perturbation in terms of local operations (e.g., spin-flip $\sigma_z$) spread away to the other sites. The growth of an operator $A(t)$ in terms of the Baker-Hausdorff-Campbell (BCH) expansion is
\begin{equation}
    A(t)= A + i t [H, A] + \frac{(i t)^2}{2!} [H, [H, A]] + \frac{(i t)^3}{3!} [H, [H, [H, A]]] + \dots\, .
\end{equation}
We see that the nested commutators of a local \( A(0) \) acting on a single site with a composite Hamiltonian that contains terms for other local sites in terms of nearest neighbor interactions result in $A(t)$ having increasing support in the many-body Hilbert space.
Assuming that at the initial time ($t = 0$), along with the operator $A$, there is another operator $B$, at a different spin-site, and it commutes with $A$. At a later time $t$, if the operator $B$ ceases to commute with $A(t)$, then it is said that the effect of the old perturbation has reached the new site.
The growth of the Hilbert-Schmidt norm of this commutator shows the degree to which the initial perturbation has affected the new perturbation at time $t$. 
The formal structure of the commutator growth as a measure of information scrambling is~\cite{devvrat_OTOC}
\begin{equation}
    \mathcal{C}_{AB}(t)={\rm Tr}\langle[A_t,B]^{\dagger}[A_t,B]\rangle.
\end{equation}
Here, $A_t$ and $B$ are arbitrary Heisenberg local operators, $\langle \cdot \rangle = {\rm Tr}(\rho \cdot)$ is the average with respect
to the thermal density matrix of the system at infinite temperature, $\rho = e^{-\beta H} /Z=\mathbf{I}/d$,
where $Z = {\rm Tr}(e^{-\beta H})$ and $\beta = 1/T$. Further, $H$ is the Hamiltonian of the system.
Many features of a quantum many-body system, along with quantum chaos, are characterized by this measure of scrambling of quantum information.
Expanding this commutator structure into its four elements, we have
\begin{align}
    \mathcal{C}_{AB}(t) = \mathcal{D}_{AB}(t) + \mathcal{I}_{AB}(t) - 2 \, \Re \{ \mathcal{F}_{AB}(t) \},
\end{align}
where $\mathcal{D}_{AB}(t) = \langle B^\dagger (A^\dagger A)_t B \rangle$, $\mathcal{I}_{AB}(t) = \langle A^\dagger{}_t (B^\dagger B) A_t \rangle$, $\mathcal{F}_{AB}(t) = \langle A^\dagger{}_t B^\dagger A_t B \rangle$.

$\mathcal{I}_{AB}(t)$ and $\mathcal{D}_{AB}(t)$ are evidently the time-ordered correlation functions whereas $\mathcal{F}_{AB}(t)$ is what defines out-of-time-ordered four-point correlation function. In this work, our purpose is to study $\mathcal{F}_{AB}(t)$, for which we coin the term $\mathcal{F}$-OTOC, used subsequently throughout the paper, to denote the distinct interferometric scheme devised for the experimental realization of OTOC.
 For the case when $A$ and $B$ are unitary operators, there is a connection between $\mathcal{C}_{AB}(t)$ and $\mathcal{F}_{AB}(t)$
\begin{equation}
    \mathcal{C}_{AB}^{Unitary}(t)=2 \left( 1 -  \Re\{ \mathcal{F}_{AB}(t) \} \right).
\end{equation}
These four-point correlators depict the decay of correlations of the initially commuting operators. In general, $\mathcal{F}_{AB}(t)$ is a complex number~\cite{Swingle_2024}, but it depends on the observables taken. Here, the observables chosen, for example, Pauli $\sigma^z$ being Hermitian, automatically lead to real $\mathcal{F}_{AB}(t)$. In chaotic quantum systems, the correlations decay quickly. Any perturbation spreads fast throughout the system, and the system loses memory of that perturbation.
To measure the $\mathcal{F}$-OTOC, $\mathcal{F}(t)=\langle A^\dagger{}_t B^\dagger A_t B \rangle$, we follow an interferometric scheme where a control qubit makes two paths accounting for different sequences of operators acting on a composite initial state of the system and the control qubit. The measurement of the control qubit after the interference process elicits the $\mathcal{F}(t)$.
In \cite{SwingleInfoScram}, such a measurement scheme was provided for a system undergoing unitary evolution, whereas in \cite{Zhang_2019}, this was modified to include the effect of GKSL evolution.
To briefly sketch both schemes, we must start with an initial system state $\ket{\psi}_S$ or $\rho_S(0)$ and a control qubit $\ket{+}_c=\frac{1}{\sqrt{2}}(\ket{0}_c+\ket{1}_c)$. Now, a sequence of unitary operations are acted on $\rho_{init}=\rho_S (0)\otimes \ket{+}\bra{+}_c$ . The sequence of unitary operators acting on this state is the following,
\begin{align}
    U_1&= \mathbf{I}_S\otimes \ket{0}\bra{0}_c + B_S \otimes \ket{1}\bra{1}_c, \nonumber \\
    U_2&=e^{-iH_S t}\otimes \mathbf{I}_c, \nonumber \\
    U_3&=A_S\otimes \mathbf{I}_c \nonumber \\
    U_4&=e^{iH_St}\otimes \mathbf{I}_c ,\nonumber \\
    U_5&= B_S \otimes \ket{0}\bra{0}_c + \mathbf{I}_S\ \otimes \ket{1}\bra{1}_c .
\end{align}
The operators $A_S$ and $B_S$ are local unitary operators in the system space. 
If $(\mathbf{I}_S \otimes \sigma^x_c)$ is measured in the final density matrix rendered by the sequence, the real part of the OTOC is obtained, as can be seen from Appendix A, below Eq.~\eqref{A11}, where the description is for a general open quantum system scenario, which subsumes the unitary case.
This protocol's ability to function with both pure and mixed states makes it easy to generalize to open quantum systems. In this work, we will adapt this formalism to general non-Markovian evolution. 

The above process incorporates both forward and backward temporal evolution. In the structure of OTOC, $\mathcal{F}(t)=\langle A^\dagger{}_t B^\dagger A_t B \rangle=\text{Tr}(A^\dagger_t B^\dagger A_t B \rho(0))=\text{Tr}( B^\dagger A_t \left(B \rho(0) \right) A^\dagger_t)$ one can observe that the order of operators hints that $B$ acts on the initial density matrix $\rho(0)$ and $A_t\cdot A_t^{\dagger}$ acts at a later time $t$, but the next $B^{\dagger}$ operator acts at time $t=0$ up to where the density matrix has to evolve in the backward direction of time. The operator $B$ acts on a different site, probing the correlation with $A(t)$, denoting the perturbation at time $t$. 
The quantity $\mathcal{F}(t)=\text{Tr}(A^\dagger_t B^\dagger A_t B \rho(0))$ shows the overlap between two states, for example, $A_t B\ket{\psi}$ and $B A_t \ket{\psi}$ for $\rho(0) =\ket{\psi}\bra{\psi}$.

For open systems, the local unitary operators in the Heisenberg picture are evolved non-unitarily under a CPTP map, denoted as $\xi(t)$. The backward evolution of the operators would mean that the system Hamiltonian $H_S$ is reversed, whereas in the total Hamiltonian $H$ for the composite unitary evolution, the bath and the interaction Hamiltonians remain the same.  In other words, if $H_f = H_S + H_E + H_{SE}$ governs forward time evolution, then $ H_b = -H_S +H_E + H_{SE} $ governs backward time evolution. The CPTP map in the forward direction of time is defined as $\xi_f(t)$.
The forward time evolution of operators is denoted  as $\left\{\xi_f^{\dagger}(t) \cdot A\right\}$ instead of $A(t)$, where $\xi_f^{\dagger}(t)$ is the adjoint map. We also denote the backward CPTP map and its adjoint map, where the driving Hamiltonian is $H_b$, as $\xi_b (t)$ and $\xi_b^{\dagger}(t)$. 
The modified protocol for open quantum systems making use of $\xi(t)$ (a general non-Markovian CPTP map, which for GKSL evolution reduces to $e^{\mathcal{L}t}$, where $\mathcal{L}$ is the Lindbladian superoperator~\cite{Zhang_2019}) is
\begin{align}\label{scheme}
    \mathcal{S}_1&= \mathcal{C}(\mathbf{I}_S\otimes \ket{0}\bra{0}_c + B_S \otimes \ket{1}\bra{1}_c), \nonumber \\
    \mathcal{S}_2&=\xi_f(t)\otimes \mathcal{I}_c, \nonumber \\
    \mathcal{S}_3&=\mathcal{C}(A_S\otimes \mathbf{I}_c) ,\nonumber \\
    \mathcal{S}_4&=\xi_b(t)\otimes \mathcal{I}_c, \nonumber \\
    \mathcal{S}_5&= \mathcal{C}(B_S \otimes \ket{0}\bra{0}_c + \mathbf{I}_S\ \otimes \ket{1}\bra{1}_c ), \nonumber \\
    \rho_f&=\mathcal{S}_5 \cdot \mathcal{S}_4 \cdot \mathcal{S}_3 \cdot \mathcal{S}_2 \cdot \mathcal{S}_1 \cdot \rho_{init}.
\end{align}
 Here, $\mathcal{I}_c$ is the identity superoperator acting on the control qubit. Since the map is obtained by tracing out the bath, the identity superoperator signifies how the control qubit remains unaffected by the action of this map. The $\mathcal{C}$ denotes the operation $\mathcal{C}(U) \cdot \rho =U\rho U^{\dagger}$. The initialized composite state $\rho_{init}$ is acted upon by the above operations in the given sequence. The operator $B_S$ acts on $\rho_S(0)$ at $t=0$; then the density matrix is evolved forward up to a time $t$ by the map $\xi_f(t)$, where the local perturbation $A_S (\cdot) A_S^{\dagger}$ is applied at a different site. Next, the system is evolved backward by the map $\xi_b(t)$ to the initial time, where $B_S^\dagger$ acts on it. Hence, the dynamics carries within itself the signature of chaotic or non-chaotic behavior. Alternatively, we apply the adjoint map of the reverse evolution at time $t$ to the operator $B_S^\dagger$, $ \left\{\xi_b^\dagger(t) \cdot B_S^\dagger\right\}$, to obtain the $B_S^\dagger{}$ operator at $t=0$. 
 Now, the OTOC is obtained by taking a trace of the final density matrix $\rho_f$ with $\sigma^x_c$ as~\cite{Zhang_2019, SwingleInfoScram}
 \begin{align}\label{openFotoc}
\mathcal{F}(t, A, B) &:= \text{Tr} \left( \sigma_c^x \, \rho_f \right) \nonumber \\
        &= \Re \, \text{Tr} \left[
            B_S^\dagger  \xi_b(t) \cdot   A_S 
           \left( \xi_f(t) \cdot \left( B_S \, \rho_S(0) \right) \right) 
           A_S^\dagger 
           \right] \nonumber \\
           &= \Re \, \text{Tr} \left[
           \left( \xi_b^\dagger(t) \cdot B_S^\dagger \right) A_S 
           \left( \xi_f(t) \cdot \left( B_S \, \rho_S(0) \right) \right) 
           A_S^\dagger 
           \right].
\end{align}
The derivation of Eq.~\eqref{openFotoc} is shown in Appendix A.

For unitary evolution, the OTOC is $\text{Tr}(B ^{\dagger} A(t)\rho(0)B A^{\dagger}(t))$. This is why it is called \textit{out-of-time ordered} correlator. A time-ordered term is $\text{Tr}(B^{\dagger} A(t)\rho(0) A^{\dagger}(t) B )$, the likes of which were discarded when tracing with respect to $\sigma^x_c$, as shown in Appendix A for general open systems. For an open system, Eq.~\eqref{openFotoc} employs a similar scheme of operations. Let us illustrate Eq.~\eqref{openFotoc} term by term: 

\begin{enumerate}
    \item $B_S$ is applied on $\rho_S(0)$ yielding $B_S\rho_S(0)$ at $t=0$. 
    
    \item Then it is time evolved by $\xi_f(t)$ up to $t$, where the spread of the perturbation is to be probed: $\xi_f(t)\cdot\left(B_S\rho_S(0)\right)$.
    
    \item At time $t$, the  perturbation $A_S(\cdot)A_S^{\dagger}$ is applied: $A_S \left(\xi_f(t)\left(B_S\rho_S(0)\right)\right)A_S^{\dagger}$. This is equivalent to the evolved perturbation $A_S(t)(\cdot)A_S^{\dagger}(t)$ being applied on $B_S\rho_S(0)$ where $A_S(t)=\xi_f^{\dagger}(t)\cdot A_S$ is obtained using the adjoint map. 
    
    \item Now taking clue from the unitary OTOC $\text{Tr}(B_S ^{\dagger} A_S(t)\rho_S(0)B_S A_S^{\dagger}(t))$, we see that $B_S^{\dagger}$ needs to come back to the initial time by a backward time evolution map $\xi_b^{\dagger}(t)$. We finally obtain $\left( \left( \xi_b^\dagger(t) \cdot B_S^\dagger \right) A_S 
           \left( \xi_f(t) \cdot \left( B_S \, \rho_S(0) \right) \right) 
           A_S^\dagger 
           \right)$.
\end{enumerate}

\subsection{Corrected $\mathcal{F}$-OTOC}
In the case of open systems, the $\mathcal{F}$-OTOC shows both the effects of information scrambling and that of open system effects. The scrambling occurs as a result of the non-commutativity between $A_S(t)$ and $B_S$ operators, and the open system effects come from the $\left( \xi_b^\dagger(t) \cdot B_S \right)$ term. To see the effects of pure scrambling, the $\mathcal{F}$-OTOC needs to be corrected by factoring out the effect of dissipation. In \cite{Zhang_2019}, a corrected OTOC was proposed by dividing $\mathcal{F}(t, A, B)$  by $\mathcal{F}(t, I, B)$ because of the fact that $\mathcal{F}(t, I, B)$ encodes purely dissipative effects as $I$ and $B$ are always commuting. The corrected $\mathcal{F}$-OTOC is defined as
\begin{align}\label{CorrectedOTOC}
    \mathcal{F}_c(t)=\frac{\mathcal{F}(t,A,B)}{\mathcal{F}(t,I,B)}.
\end{align}

\section{Variation of the $\mathcal{F}$-OTOC and corrected $\mathcal{F}$-OTOC, in open system models}\label{nmgad}
Here we calculate the  $\mathcal{F}$-OTOC and corrected $\mathcal{F}$-OTOC for a light-matter interaction model (TC) and a spin chain interaction model (TFIM). The non-Markovian nature of these models is briefly discussed in Appendix B. These models offer a variety of scenarios to study information scrambling. They differ in the type of interaction with the bath, viz., local or global interaction. In every scenario, the information scrambles either within the system or to the bath. We analyze both the perspectives and the degree to which the nature of coupling to the bath affects the scrambling.
\subsection{Tavis Cummings model}
Here, we consider the $N$-qubit TC model consisting of $N$ two-level atoms with transition frequencies $\omega_0$, coupled to a single mode of a quantized radiation field of frequency $\omega_c$~\cite{RH_Dicke, Kirton_review}. The Hamiltonian ($\hbar = 1$) of the system is given by
\begin{align}
    H^{\text{TC}}& = H^{TC}_S+H^{TC}_E+H_{SE}^{TC} \nonumber \\ 
    H^{TC}_S&=\omega_0  \sum_{i=1}^4 \sigma^z_i + j_s\sum_{i=1}^{3} \sigma^z_i \sigma^z_{i+1}\nonumber \\
    H^{TC}_E&=\omega_c a ^\dagger a \nonumber \\
    H_{SE}^{TC}&=\frac{\lambda}{2\sqrt{N}} \sum_{i=1}^4 \left( \sigma^+_i a +  \sigma^-_i a^\dagger \right),
\end{align}
where $\sigma^k$ for ($k=x, y$, and $z$) are the Pauli spin matrices and $a$ ($a^\dagger$) are the annihilation (creation) operators. Here, we take $N = 4$ throughout the paper. Further, the initial states of the system and the bath are taken to be $\rho_S(0) = (\ket{\psi(0)}_S\bra{\psi(0)}_S)^{\otimes N}$, where $\ket{\psi_S(0)} = \frac{\sqrt{3}}{2}\ket{0} + \frac{1}{2}\ket{1}$, and $e^{-H^{TC}_E/T}/ {\rm Tr}\left[e^{-H^{TC}_E/T}\right]$, respectively, where $T$ is the temperature of the bath. The initial states are chosen arbitrarily, as the $\mathcal{F}$-OTOC being an operator-dependent measure is not significantly affected by the choice of initial state. This has also been checked numerically. The $H^{\rm TC}_S$ of the TC model corresponds to the $\theta = 0$ case of the TFIM, see Eq.~\eqref{Ham_Ising} below, and is trivially integrable.
\begin{figure}
    \centering
    \includegraphics[width=1\linewidth]{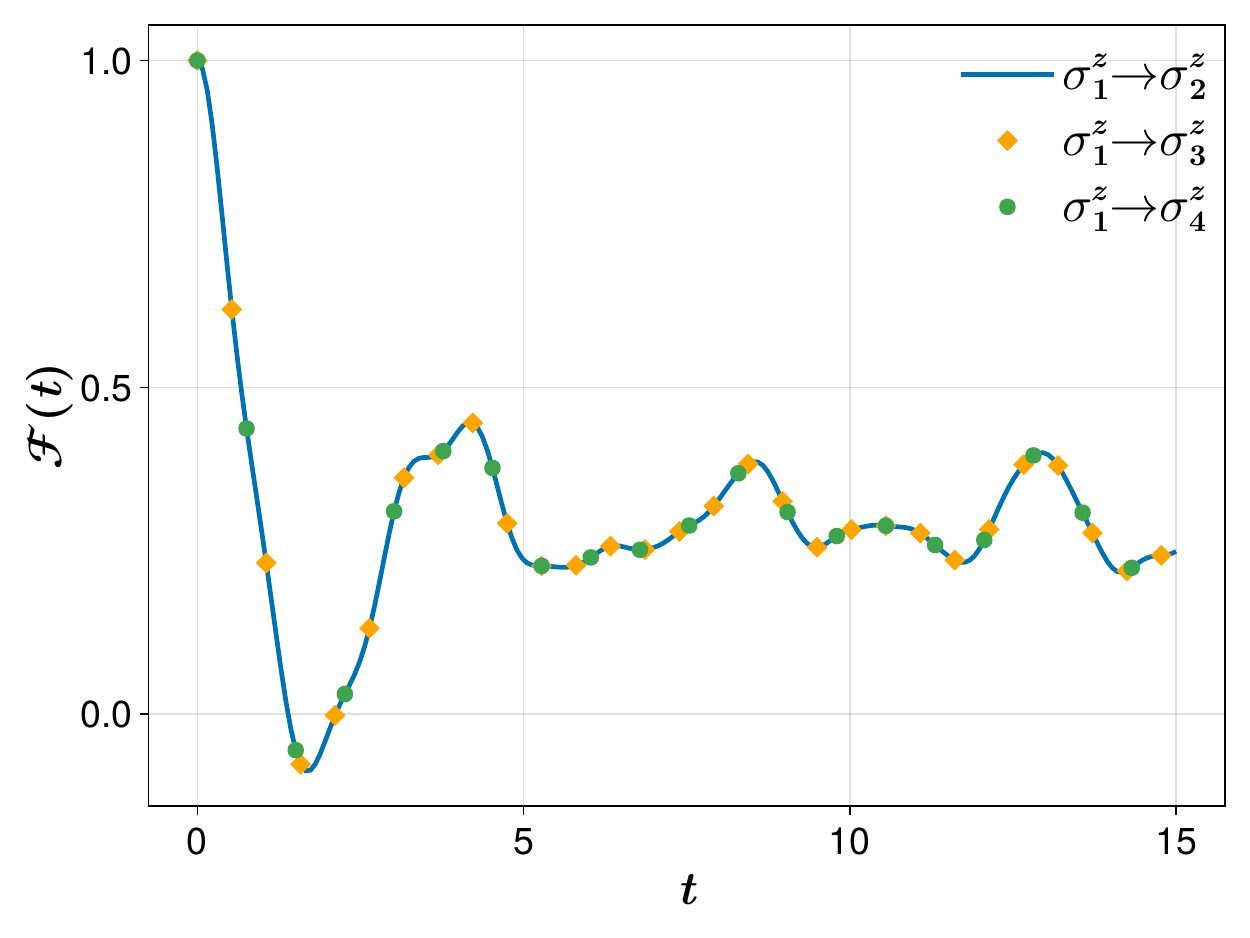}
    \caption{The plots are of the $\mathcal{F}$-OTOC $\mathcal{F}(t)$ for the Tavis Cummings model for $j_s=0$. The other parameters are, $j_{TC}=\frac{\lambda}{2\sqrt{N}}=0.5$, $\omega_0=2,~\omega_c=2,~T=10$.}
    \label{fig:Fotoc_TC_js0}
\end{figure}

\begin{figure}
    \centering
    \includegraphics[width=1\columnwidth]{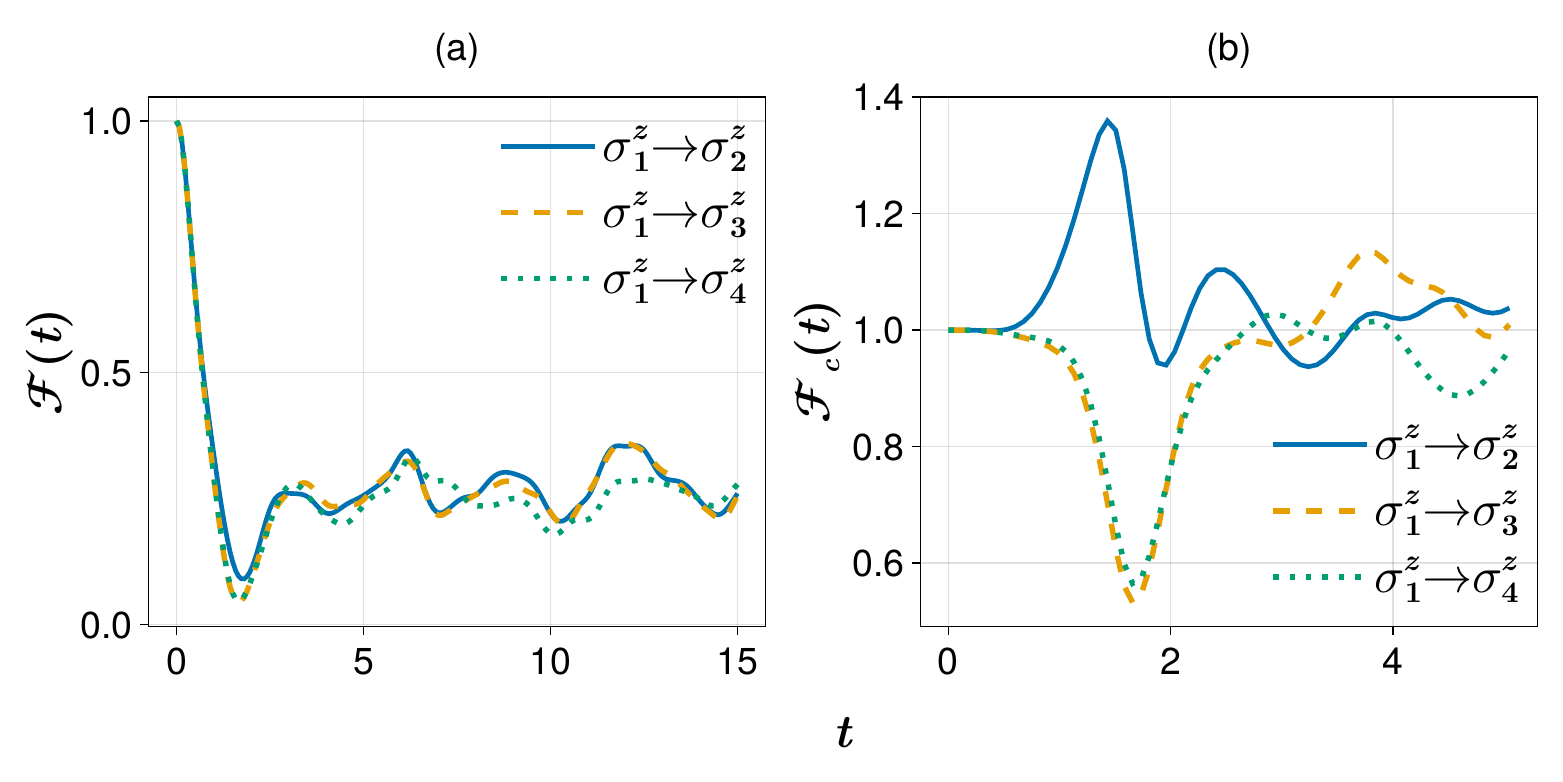}
    \caption{The plots are of the $\mathcal{F}$-OTOC $\mathcal{F}(t)$  in (a) and the corrected $\mathcal{F}$-OTOC $\mathcal{F}_c(t)$ in (b) for the Tavis Cummings model for $j_{TC}=\frac{\lambda}{2\sqrt{N}}=0.5$ and the interaction strength $j_s=0.5$  of the nearest neighbor interaction in the system spin chain. The other parameters are $\omega_0=2,~\omega_c=2,~T=10$.}
    \label{fig:Fotoc_Cor_TC_AB}
\end{figure}

In all the figures, the right arrow ($\rightarrow$) denotes the direction of scrambling from the site of $B_S$ to $A_S$. In the structure of OTOC, we see that the operator $B_S$ acts on the initial density matrix, perturbing the spin at the site. This perturbed probe site evolves in time, and the scrambling occurs in the state space up to the time $t$. At $t$, $A_S$ is applied, and the backward journey is initiated, after which $B_S$ acts at $t=0$.

In the TC model, for $j_s=0$, i.e., when there is no interaction among the spins, we see in Fig.~\ref{fig:Fotoc_TC_js0} that the choice of $A_S$ and $B_S$ doesn't matter and scrambling is the same for any two local operators. The decaying profile of $\mathcal{F}$-OTOC shows the effect of the bath, which acts individually on each of the non-interacting four spins.

For this model at $j_s\neq0$, the $\mathcal{F}$-OTOC is shown in Fig.~\ref{fig:Fotoc_Cor_TC_AB}(a). It can be observed from the plot of $\mathcal{F}(t)$ that in the presence of dissipation, the initial correlations drop drastically irrespective of the distance at which the perturbation is measured at time $t$. This is because each spin is attached to a non-Markovian bath, and the information scrambling from site to site is suppressed by the rapid dissipation of the operators. Due to the non-Markovian nature of the bath, the $\mathcal{F}(t)$ exhibits revivals after the initial drastic decay.

In Fig.~\ref{fig:Fotoc_Cor_TC_AB}(b), the corrected $\mathcal{F}$-OTOC, $\mathcal{F}_c(t)$, is depicted, and it is observed that the light cone is partially retrieved. Light cone refers to the scenario when $\mathcal{F}(t)$ maintains its correlation (i.e., $\mathcal{F}(t)=1$) for a longer duration when scrambling is observed at a farther spin. In Fig.~\ref{fig:Fotoc_Cor_TC_AB}(a), this is not observed because of dissipation. This is why $\mathcal{F}_c(t)$ is used, where the light cone is recovered since the effect of dissipation is suppressed. One can note that the value of $\mathcal{F}_c(t)$ can go higher than one since it is a ratio of two $\mathcal{F}(t)$'s. Therefore, it need not follow the constraint on $\mathcal{F}(t)$ to be less than one. 
In Fig.~\ref{fig:Fotoc_TC_js0}, the initial sharp decay is attributed to the dissipation of information to the radiation bath that is connected to each and every spin of the system. This strong effect is present in both the interacting and the non-interacting cases and is predominant. In the case of the four-spin chain having $\sigma^z-\sigma^z$ type interaction with strength $j_s$, despite it being an integrable model, corresponding $\mathcal{F}$-OTOC profiles emerge as those of a chaotic system because of the presence of strong dissipation effects. In Fig.~\ref{fig:Fotoc_Cor_TC_AB}(a), the mild differences between the $\mathcal{F}$-OTOCs of different transitions are caused by the information being scrambled to other spins due to interaction between them.

The corrected $\mathcal{F}-$OTOC [$\mathcal{F}_c(t)$] is defined in such a way that it omits the dissipation of the $B_S$ operator and not that of the $A_S$. Consequently, it can be expected that the $\mathcal{F}_c(t)$ can only partially retrieve the light cone structure. In the light cone in Fig.~\ref{fig:Fotoc_Cor_TC_AB}(b) corresponding to the TC model, the three $\mathcal{F}$-OTOCs drop in times that are not equidistant. The light cones in $\mathcal{F}$-OTOCs for TFIM models, Fig.~\ref{fig:FOTOC_for_TFIM_two_angles}, have equidistant drops of $\mathcal{F}$-OTOCs for different transitions since the system is only locally connected to the bath and suffers minor dissipation. The system of the TC model is an integrable model and is similar to the system Hamiltonian of the TFIM model at $\theta=0$. It has been checked that the $\mathcal{F}$-OTOC for the $\theta=0$ case remains one for all times, which means information is not scrambled at all. This suggests that if one gets rid of the dissipation effect from the TC model, there should not be any scrambling such that $\mathcal{F}(t)=1$. We don't see that in Fig.~\ref{fig:Fotoc_Cor_TC_AB}(b). The flawed light cone in Fig.~\ref{fig:Fotoc_Cor_TC_AB}(b) can thus be said to have been partially retrieved. Also, the initial rise and drop of the $\mathcal{F}_c(t)$ can be attributed to the interaction with the bath, while the revivals leading to erratic oscillations can be attributed to the smallness of the number of bath spins and the non-Markovian effects of the bath.
The featureless oscillations that appear after the initial decay arise because of the limited number of spins in the system. For a chaotic many-body system with a number of spins in the thermodynamic limit, one expects attainment of a thermal saturation value after the initial decay~\cite{García-Mata:2023}. In addition to that, our models are constructed with an arbitrary coupling constant and hence are eligible for non-Markovian behavior. The featureless oscillations after the initial decay can be attributed to either of the above two reasons.

\subsection{Tilted field Ising model}\label{tfim}
We consider an Ising chain, which interacts with a tilted magnetic field. This model has been found to be useful for understanding the chaotic behavior of a quantum system and has been studied extensively from statistical physics and quantum information perspectives~\cite{TiltedMfieldofarul}. The Hamiltonian $H_{\theta}$ for this model is given by
\begin{align}
    H_{\theta}(J, \mathcal{B},\theta)= H_S+H_E+H_{SE},
\end{align}
    where,
\begin{align}\label{Ham_Ising}
    H_S &= \mathcal{B} \sum_{i=1}^4 \left(\sin(\theta) \sigma^x_i + \cos(\theta) \sigma^z_i\right)+J\sum_{i=1}^{3} \sigma_{i}^z \sigma_{i+1}^z, \nonumber \\
    H_E &= \mathcal{B} \sum_{i=5}^8 \left(\sin(\theta) \sigma^x_i + \cos(\theta) \sigma^z_i\right)+J\sum_{i=5}^{7} \sigma_{i}^z \sigma_{i+1}^z, ~~~~\text{and} \nonumber \\
    H_{SE} &=J \sigma_{4}^z \sigma_{5}^z.
\end{align}
The TFIM (Tilted field Ising model) is a system of 4 spins Ising chain with nearest neighbor interaction, interacting with another 4 spins Ising chain with nearest neighbor interaction, considered as a bath. The interaction between the system and bath happens at the edge of both chains, i.e., between $\sigma_4$ and $\sigma_5$. Compared to the TC model, where the bath is attached to each one of the spins, for TFIM, the dissipation effects are milder.
\begin{figure}
    \centering
    \includegraphics[width=1\linewidth]{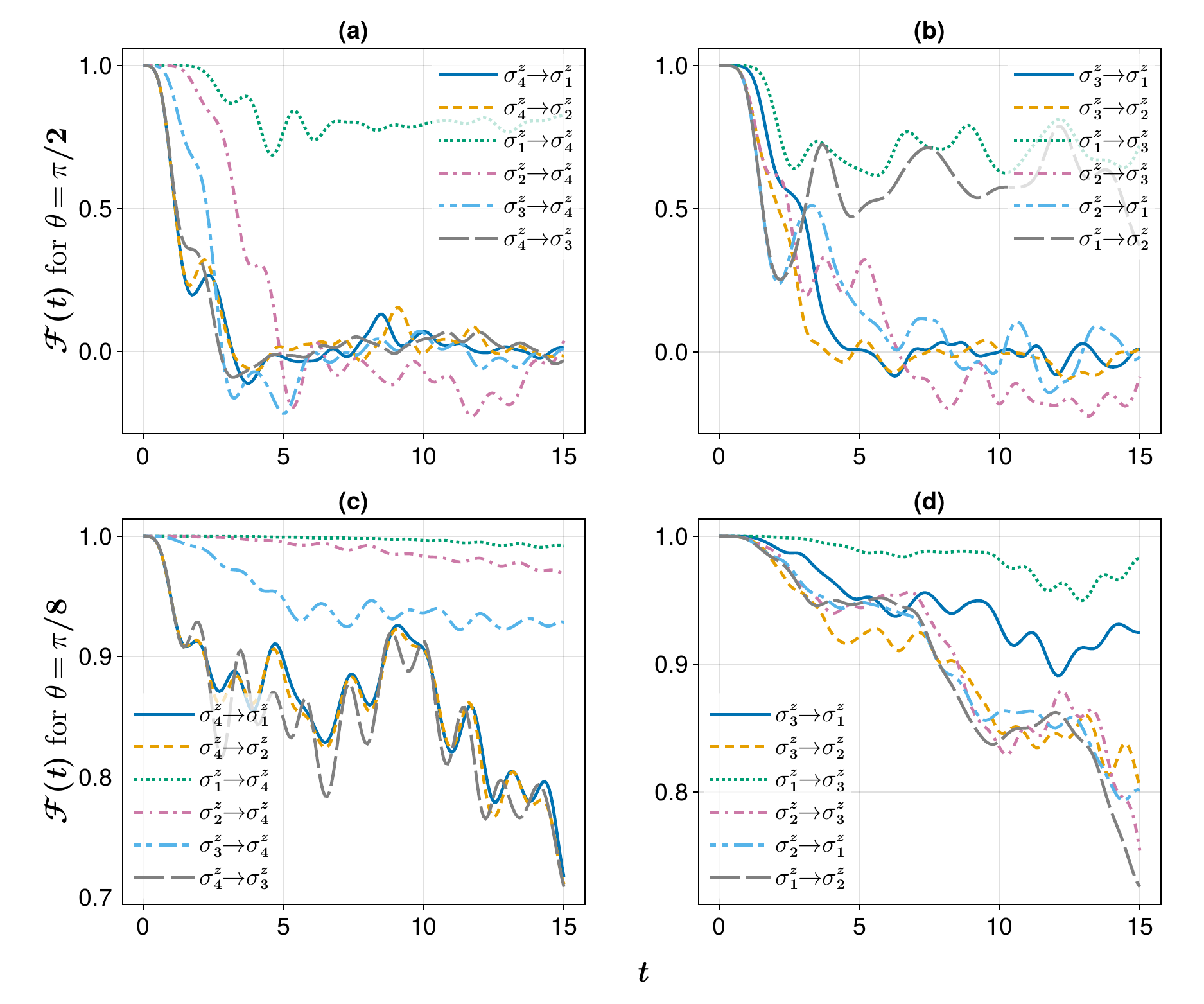}
    \caption{The $\mathcal{F}$-OTOC for the TFIM model of four system spins interacting with a four Ising spin bath. The direction of the scrambling shown in the figures by means of an arrow denotes that the information of the action of $B_S(0)$ scrambles up to time $t$ where $A_S$ is applied. In (a) and (b) the $\mathcal{F}$-OTOC is plotted for 
 $\theta=\pi/2$. For (c) and (d), the angle is $\theta=\pi/8$. The other parameters chosen are $\mathcal{B}=0.5$, $j=0.8$.} 
    \label{fig:FOTOC_for_TFIM_two_angles}
\end{figure}
Here, we take the initial states of the system and bath to be $\rho_S(0) = (\ket{\psi(0)}_S\bra{\psi(0)}_S)^{\otimes 4}$, and $\rho_E(0) = (\ket{\psi(0)}_E\bra{\psi(0)}_E)^{\otimes 4}$, respectively, where $\ket{\psi(0)}_{S(E)}= \frac{\sqrt{3}}{2}\ket{0} + \frac{1}{2}\ket{1}$.
At $\theta=0$, $\mathcal{F}$(t) is 1. This means that the initial correlation of the commuting operators remains intact for this integrable case. It is to be noted that for $\theta=0$, the TFIM's system matches that of the TC model's system, which is an integrable Hamiltonian. The $\mathcal{F}$-OTOC for the TC model, even though integrable, decays because of the action of the bath on each of the spins in the system.
In Fig.~\ref{fig:FOTOC_for_TFIM_two_angles}, the $\mathcal{F}$-OTOC for TFIM model is plotted. For $\theta=\pi/2$, which is a non-trivially integrable transverse magnetic field model, we observe that the scrambling of information is less when $B_S$ is $\sigma^z_1$ because the decay in $\mathcal{F}(t)$ is lower as compared to the case when $B_S$ is measured at other spin sites.
Since the model is such that spin 1 is connected to only spin 2, whereas all the other system spins are connected to spins on both sides, a lesser amount of scrambling from spin 1 is justified.  This also explains the difference in the behavior of $\mathcal{F}$-OTOC for the transport from spin 1 to spin 3 and that of spin 3 to spin 1. Scrambling from spin 1 is lower since it has only one neighbor, whereas spin 3, having two neighbors, scrambles faster despite the information being traveled through the same distance. 

In this model, the system spins are not connected to a bath except for the 4th spin, which is connected to the first spin of the bath spin chain, i.e., the 5th spin. This is why the open system effects of dissipation of the system operators are not immediate like those of the TC model.
In the cases for which $B_S$ is the local operator at the 4th site, connected to the bath spin 5, the light cone is not observed for the transfer of information to the 3rd, 2nd, and 1st sites. In all three cases, $\mathcal{F}(t)$ decays immediately despite traveling to spin sites at different distances.

For all the other cases when $B_S$ is not $\sigma_z^1$ or the bath-adjacent $\sigma_z^4$, we observe the generic behavior of the light cone where the initial pause at $\mathcal{F}(t)\simeq1$ is more for greater distances between the spins. When the angle of tilt $\theta=\pi/8$, for which the model is non-integrable, we see almost all the properties mentioned above but with a lesser degree of clarity, which could be ascribed to the non-integrability of the model. The distinction of the behavior of the spin 1 seen at TFIM for $\theta=\pi/2$ is lost in the case of the TFI model at $\theta=\pi/8$, see Figs.~\ref{fig:FOTOC_for_TFIM_two_angles}(c) and (d).
That is why the integrability of the TFI model at $\theta=\pi/2$ works as a probe for the fact that spin 1 scrambles less information through its only connection to spin 2, but all other spins scramble more through their connection to spins on both sides. The TFIM at $\theta=\pi/2$ also exhibits a special behavior of the spin at the 4th site, as it is the only spin connected to the bath spin chain and is responsible for the dissipation effects. This distinction is also lost when $\mathcal{F}(t)$ is observed for the non-integrable TFIM at $\theta=\pi/8$.

From the above analysis, it could be remarked that since only the TFIM for angles $\theta>0$ exhibit fast and pure scrambling of information, these Hamiltonians are chaotic.

\begin{figure}
    \centering
    \includegraphics[width=1\linewidth]{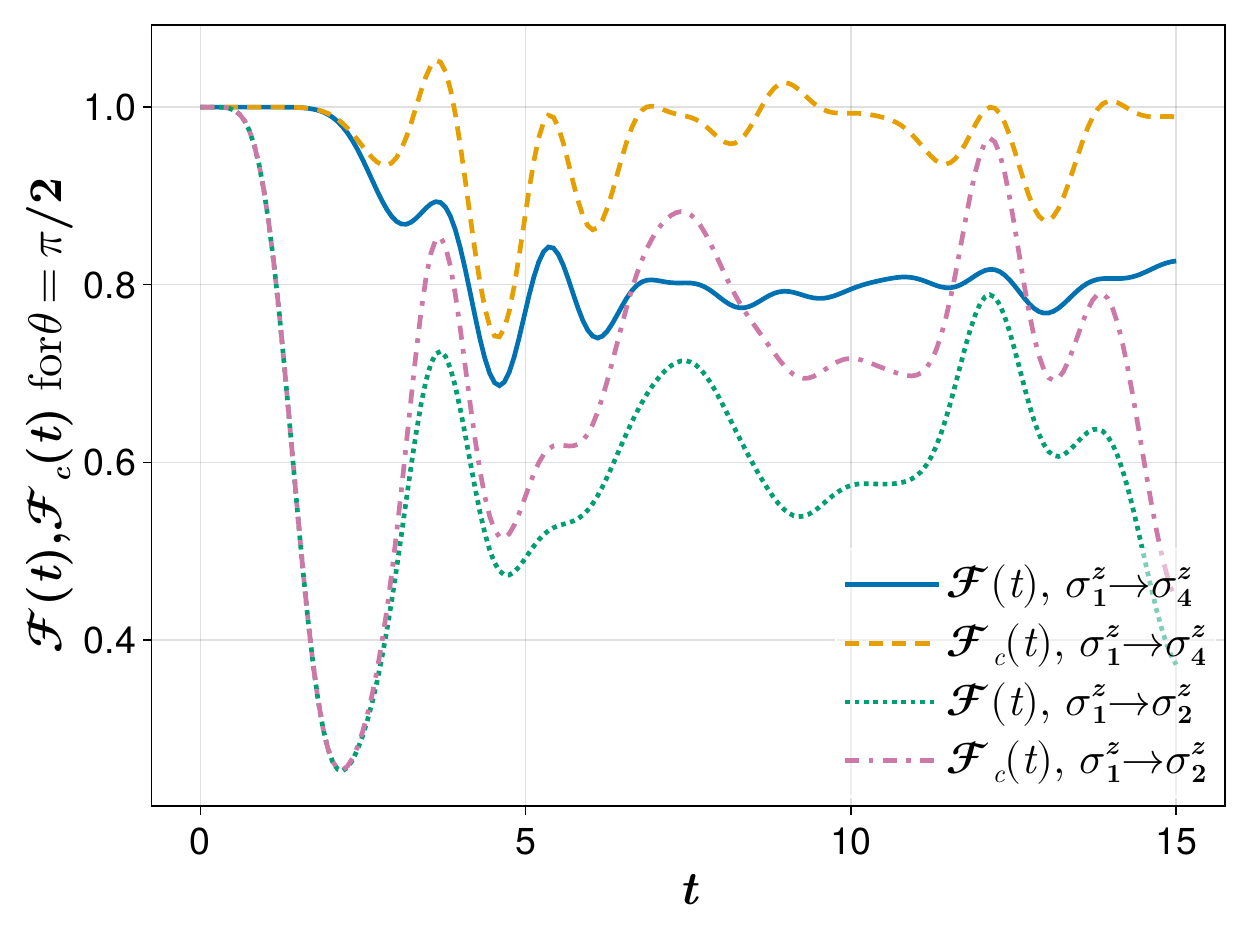}
    \caption{The corrected OTOC $\mathcal{F}_c(t)$ without dissipation effects and the $\mathcal{F}(t)$ with dissipation effects are plotted for the TFIM model at $\theta=\pi/2$. The other parameters chosen are $\mathcal{B}=0.5$, $j=0.8$.}
    \label{fig:Cor_Fotoc_pi2_Compare}
\end{figure}

\begin{figure}
    \centering
    \includegraphics[width=1\linewidth]{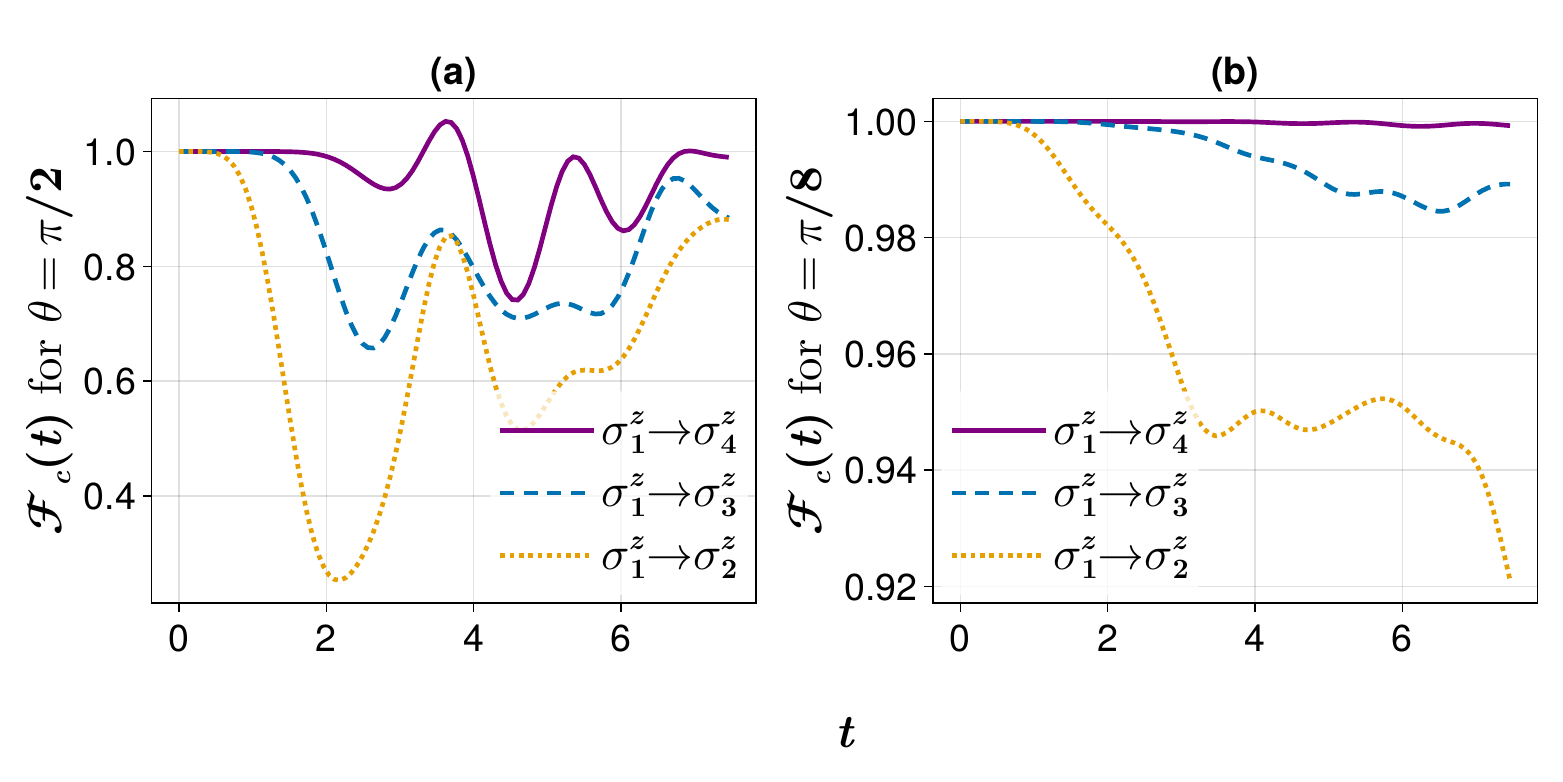}
    \caption{The corrected $\mathcal{F}$-OTOC for the TFIM model of four system spins interacting with a four Ising spin bath. The direction of the scrambling shown in the figures by means of an arrow denotes that the information of the action of $B_S(0)$ scrambles up to time $t$ where $A_S$ is applied. In (a) and (b) the corrected $\mathcal{F}$-OTOC, $\mathcal{F}_c(t)$ is plotted for $\theta=\pi/2$ and $\theta=\pi/8$ respectively. The other parameters chosen are $\mathcal{B}=0.5$, $j=0.8$.}
    \label{fig:Cor_FOTOC_TFIM}
\end{figure}

To illustrate the impact of dissipation on the information scrambling in the TFIM, we plot $\mathcal{F}_c(t)$ and $\mathcal{F}(t)$ for $\theta=\pi/2$ in Fig.~\ref{fig:Cor_Fotoc_pi2_Compare}. Interestingly, we observe that decay in both the $\mathcal{F}(t)$ and the $\mathcal{F}_c(t)$ occurs at similar times, denoting information scrambling. However, the values of $\mathcal{F}_c(t)$ after the initial drop always remain above or equal to the values of $\mathcal{F}(t)$. This brings out the light cone structure for both $\mathcal{F}_c(t)$ and $\mathcal{F}(t)$.
In Fig.~\ref{fig:Cor_FOTOC_TFIM}, we depict $\mathcal{F}_c(t)$ for the TFIM model for $\theta = \pi/2$ and $\theta = \pi/8$. 
In this figure, we observe that it takes a longer duration for $\mathcal{F}_c(t)$ to start decaying when $B_S$ and $A_S$ are measured at more distant spin sites. This indicates that the effect of information scrambling takes longer to propagate to farther spins. The time at which scrambling starts increases as we step away from the initial spin 1. This again confirms the existence of an information light cone. An information light cone is proof that information travels ballistically through the spins at a definite speed. This speed is defined as the Lieb-Robinson velocity ~\cite{LiebRobSwingle_PhysRevLett.117.091602}. In~\cite{Zhang_2019} as well, it is shown that for the dissipation of GKSL nature, $\mathcal{F}(t)$ has its light cone destroyed, whereas the $\mathcal{F}_c(t)$ partially revives the light cone. For all cases, we see that the $\mathcal{F}(t)$ decays but does not saturate to any specific value, and fluctuations are seen. This could be attributed to the non-Markovianity of the models considered here.
We have also calculated the $\mathcal{F}(t)$ using $B_S=\sigma_z$ and $A_S=\sigma_x$. The general trend of the results shown and discussed above remains the same for this choice.

Generally, $\mathcal{F}$-OTOC starts decaying from one, and specifically for chaotic systems, they undergo a drastic initial decay, denoting fast scrambling of information. In Fig.~\ref{fig:FOTOC_for_TFIM_two_angles}, we observe this general behavior being replicated to various degrees of similarity except for the $\mathcal{F}$-OTOC for TFIM at $\theta=\pi/8$. It doesn't show the drastic initial decay characteristic of OTOCs for chaotic many-body systems for all sets of initial and final spins. For $\theta=0$, the $\mathcal{F}$-OTOC is one for all possible transitions, and for $\theta=\pi/2$, despite being non-trivially integrable, the $\mathcal{F}$-OTOC exhibits behavior that very closely corresponds to that of a chaotic system. We have seen in our numerical calculations that for the intermediate angles, $\mathcal{F}$-OTOC also shows behavior that is intermediate between the two extremes at $\theta=0$ and $\theta=\pi/2$. From this, we can infer that for the intermediary angles, the open system TFIM we have considered is not highly chaotic, as it does not show fast scrambling for all possible transitions.

\section{Loschmidt Echo}\label{loshcmidt echo}
Loschmidt Echo quantifies the irreversible behavior of a quantum system by analyzing the effect of a perturbation in the Hamiltonian of the system. The above discussions of $\mathcal{F}$-OTOC~\cite{Zhang_2019} use a scheme where the forward and backward evolution of one of the operators $A_S(t)$ is done by two different Hamiltonians. Keeping the bath and the interaction Hamiltonian intact, the system Hamiltonian $H_S$ is reversed for the backward evolution. We define Loschmidt echo in a way such that an initial separable composite state $\rho_S(0)\otimes\rho_E(0)$ is made to evolve forward in time by the Hamiltonian $H_f=H_S+H_E+H_{SE}$ up to  $t$ when it is made to come back to the initial time by a backward evolution with the Hamiltonian $H_b=-H_S+H_E+H_{SE}$. The distance between the initial and the final system states at the same point in time is measured in terms of the overlap of two density matrices. If no perturbation $\Delta$ in the Hamiltonian of the backward evolution is added, the Loschmidt echo decays, illustrating the dissipation effects on the evolution. We put $\Delta$ as an additive perturbation to $H_S$ in the backward evolution and see its effect on the decaying Loschmidt echo. Here, we have taken the above-mentioned models to calculate the Loschmidt Echo.
Consider the `Loschmidt echo' as a two-point correlation function that illustrates the notion of an `echo' for an open system, 
\begin{align}
    \mathcal{L}_E(t)&={\rm Tr}\left[A_S \xi_b \cdot \xi_f \cdot(  B_S \rho_S(0))\right].
\end{align}
An operator $B_S$ is applied to the initial state $\rho_S(0)$, and the state is made to evolve forward and backward in time with slightly different system Hamiltonians. Again, at $t=0$, operator $A_S$ is applied to calculate the two-point correlation function. If $A_S=\rho_S(0)$ and $B_S=\mathbf{I}$, as considered here, then we have the Loschcmidt echo
\begin{align}
    \mathcal{L}_E(t)&={\rm Tr}\left[\rho_S(0) \rho'_S(0)\right],
\end{align}
where
\begin{align}
    \rho'_S(0)&=\xi_b \cdot \xi_f \cdot \rho_s(0) \nonumber \\
    &={\rm Tr}_B \left( e^{-iH_b t}e^{-iH_f t} \left( \rho_S(0) \otimes \rho_E(0) \right) e^{iH_f t} e^{i H_b t} \right).
\end{align}
Here, the $\rho_S'(0)$ is the state after the backward evolution of $\rho_S(t)$ by the system Hamiltonian $-(H_S+\Delta)$.
We observe the effect of dissipation and the perturbation to the Hamiltonian $\Delta$ on an initial uncorrelated state using the Loschmidt Echo for a light-matter interaction model (TC model) and an Ising spin chain model (TFIM).
\subsection{Tavis Cummings model}
We observe for the Tavis Cummings model, how the Loschmidt Echo varies with the given perturbation $\Delta=\omega_d\sigma^z$, for different values of $\omega_d$. The coupling parameter of the system and environment is kept constant, $j_{TC}=\lambda/(2\sqrt{N})=0.5$. The composite Hamiltonian for the backward evolution is defined as $H_b=-(H_S+\Delta)+H_E+H_{SE}$.
\begin{figure}
    \centering
    \includegraphics[width=1\linewidth]{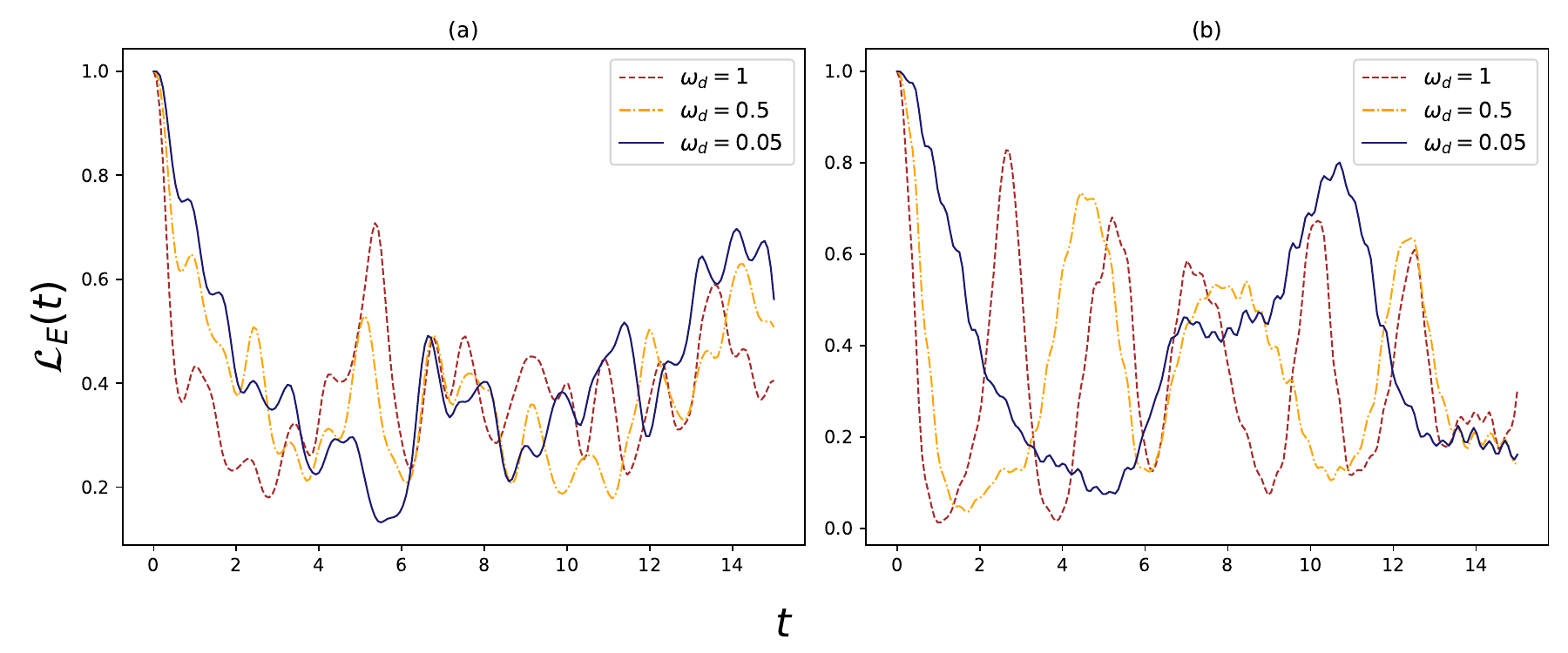}
    \caption{The evolution of the Loschmidt echo for the TC model. The three figures in each of the plots are for different strengths of perturbation $\Delta = \omega_d\sigma^z$. In (a), $\omega_0=2,~\omega_c=2$ and for (b) $\omega_0=2,~\omega_c=8$. Other parameters are the coupling strength $j_{TC}=\frac{\lambda}{2\sqrt{N}}=0.5$ and the nearest neighbor interaction strength $j_s=0.5$ in the system spin chain. }
    \label{fig:LEcho_TC_vary_wd}
\end{figure}

In Fig.~\ref{fig:LEcho_TC_vary_wd}, we see the decay of the Loschmidt echo for a small perturbation $\Delta$ in the Hamiltonian of the backward evolution. The TC model is integrable. The decay of $\mathcal{L}_E(t)$ means that the final state is different from the initial state. For an integrable Hamiltonian system, this could be attributed to the presence of the bath. When different perturbations are added, we see the decaying nature of $\mathcal{L}_E(t)$ is sustained. For $\omega_0=\omega_c = 2$, there is not much of a qualitative difference among the plots for different values of $\Delta=\omega_d\sigma_z$, except that the early time decay is steeper for the larger value of $\omega_d$. When $\omega_0=2$ and $\omega_c=8$, we observe a difference in the decaying profiles of $\mathcal{L}_E(t)$. With larger values of perturbation, the decay of $\mathcal{L}_E(t)$ becomes increasingly oscillatory in nature. In this case, also, the steepness of the initial decay is higher for larger values of $\Delta$. In the scenario of a sluggish bath~\cite{Breuer2007}, which could be obtained when the frequency of the bath is much higher than that of the system, the system acts almost like a free entity. That is why, in the second case, the perturbation being added acts to decrease the effect of the bath, and the integrability of the TC model comes back as evident from the oscillatory nature of the Loschmidt echo.

\subsection{Tilted field Ising model}
For TFIM model, we plot the $\mathcal{L}_E(t)$ at angles for which the $H_{\theta}$ is trivially integrable ($\theta=0$), non-integrable ($\theta = \pi/8, 7\pi/16$), and non-trivially integrable ($\theta=\pi/2$).
\begin{figure}
    \centering
    \includegraphics[width=1\linewidth]{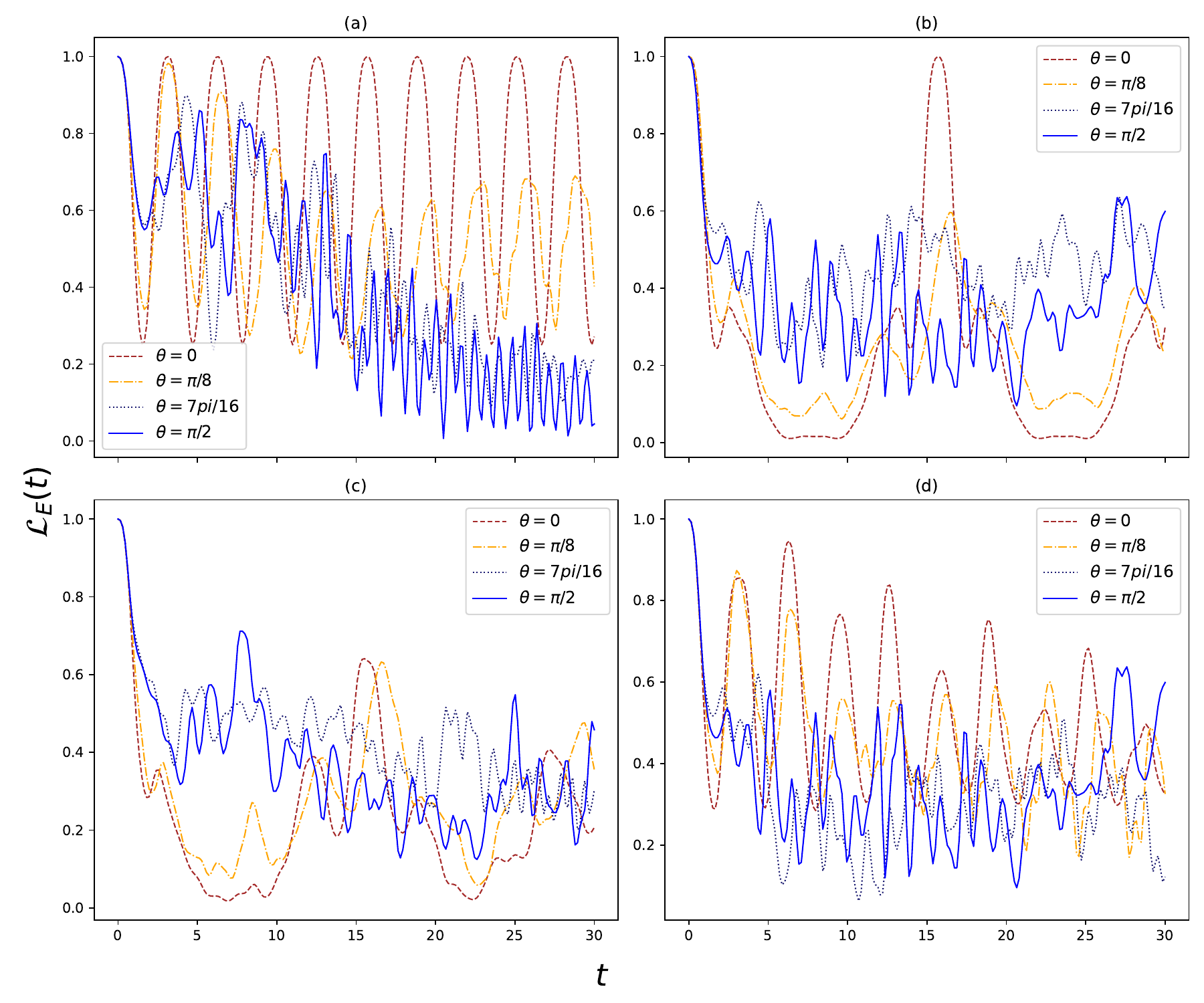}
    \caption{The variation of $\mathcal{L}_E(t)$ for TFIM model. For (a), the perturbation $\Delta$ is zero, and for (b), it is $\Delta_1$, see Eq.~\eqref{Deltas}. For (c) and (d), it is $\Delta_2$ and $\Delta_3$, respectively. The magnitude of the perturbation is $w_d=0.2$. The other parameters chosen are $\mathcal{B}=J=0.5$. }
    \label{fig:LEchoU_TFIM_vary_pert}
\end{figure}
In Fig.~\ref{fig:LEchoU_TFIM_vary_pert}(a) we see for zero perturbation the $\mathcal{L}_E(t)$ has a perfect oscillatory nature for the $H_\theta$ at $\theta=0$ with complete revivals at a certain interval attesting to the fact that for $\theta=0$ the TFI model is integrable, non-chaotic and therefore does not show scrambling. As $\theta$ is increased to values ($\pi/8$,$7\pi/16$) for which $H_\theta$ is non-integrable, the periodic nature of the oscillation of $\mathcal{L}_E(t)$ breaks, and it starts decaying, hinting at the chaotic nature of non-integrable Hamiltonians along with the dissipation effects of the bath. At $H_\theta$'s non-trivially integrable limit, $\theta=\pi/2$, however, the revival of the oscillatory nature is not observed. This leads us to infer that, as a state-dependent measure of information scrambling, viz., the Loschmidt Echo, doesn't probe the non-trivial integrability that the operator-dependent measures of information scrambling are capable of. Thus, for example, in operator-dependent measures such as $\mathcal{F}$-OTOC or the Bipartite OTOC~\cite{zanardibipotoc2, Bose_2024}, the dissimilarity between non-trivially integrable and non-integrable models was apparent. 

We use three kinds of perturbations to see how they affect the dissipative nature of the Loschmidt echo in open systems. The system Hamiltonian of TFIM is $ H_S = \mathcal{B} \sum_{i=1}^4 \left(\sin(\theta) \sigma^x_i + \cos(\theta) \sigma^z_i\right)+J\sum_{i=1}^{3} \sigma_{i}^z \sigma_{i+1}^z $. Here $\mathcal{B}=J=0.5.$ The added perturbations have magnitude $w_d=0.2 ~(<\mathcal{B}=0.5)$. The three kinds of perturbations are:
\begin{align}\label{Deltas}
    \Delta_1&=\omega_d \sum_{i=1}^4 \left(\sin(\theta) \sigma^x_i + \cos(\theta) \sigma^z_i\right),  \nonumber \\
    \Delta_2&=\omega_d \sum_{i=1}^4 \left( \sigma^x_i +  \sigma^z_i\right),  \nonumber \\
    \Delta_3&=\omega_d \sum_{i=1}^4 \sigma^x_i.
\end{align}
Consider the $\theta = 0$ case, that is, the trivially integrable Ising model. In Fig.~\ref{fig:LEchoU_TFIM_vary_pert}(b), with the perturbation $\Delta_1$, the $\mathcal{L}_E(t)$ is still oscillatory but with a very deformed wave profile compared to that of Fig.~\ref{fig:LEchoU_TFIM_vary_pert}(a). In  Fig.~\ref{fig:LEchoU_TFIM_vary_pert}(c), we see for $\Delta_2$, the oscillatory behavior breaks and $\mathcal{L}_E(t)$ starts decaying. For $\Delta_3$, decay of $\mathcal{L}_E(t)$ is again observed, and it looks like the deformed version of $\mathcal{L}_E(t)$ for the unperturbed Hamiltonian in Fig.~\ref{fig:LEchoU_TFIM_vary_pert}(a).
When perturbations $\Delta_2$ and $\Delta_3$ are added to the Hamiltonian of the backward evolution, we see the collapse of the oscillatory nature of $\mathcal{L}_E(t)$. This is because, in these two cases, the involvement of the $\sigma_x$ term makes the backward Hamiltonian non-integrable. 

\begin{figure}
    \centering
    \includegraphics[width=1\linewidth]{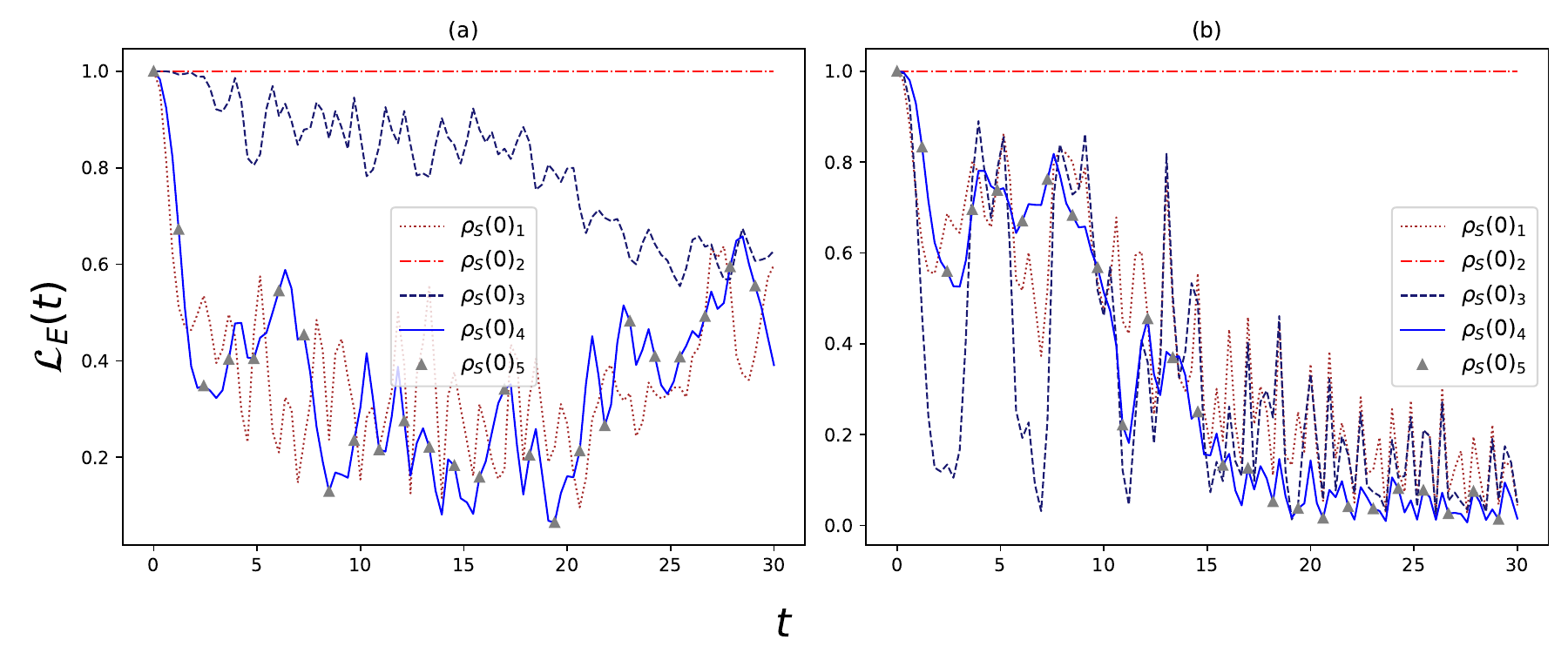}
    \caption{Loschmidt echo for TFIM model at $\theta=\pi/2$ for different initial states. In (a), the perturbation to the Hamiltonian is $\omega_d=0$, and in (b) $\omega_d=0.2$. The other parameters are $\mathcal{B}=J=0.5$}
    \label{fig:LEchoU_TFIM_varyRhoS0}
\end{figure}

The Loschmidt echo is a state-dependent measure and is influenced by the initial state of the system.
In Fig.~\ref{fig:LEchoU_TFIM_varyRhoS0} we plot the Loschmidt echo for the TFIM model in its chaotic regime ($\theta=\pi/2$) for different initial states. 
The following initial states are considered for the system spins
\begin{align}\label{eq_initial_states}
    \rho_S(0)_1&=\left[\left(\frac{\sqrt{3}}{2}\ket{0}+ \frac{1}{2}\ket{1}\right) \left(\frac{\sqrt{3}}{2}\bra{0}+ \frac{1}{2}\bra{1}\right)\right]^{\otimes 4},\nonumber \\
    \rho_S(0)_2&=\left(\frac{\mathbb{I}_2}{2}\right)^{\otimes 4},\nonumber \\
    \rho_S(0)_3&=\left[\left(\frac{1}{\sqrt{2}}\ket{0}+ \frac{1}{\sqrt{2}}\ket{1}\right) \left(\frac{1}{\sqrt{2}}\bra{0}+ \frac{1}{\sqrt{2}}\bra{1}\right)\right]^{\otimes 4},\nonumber \\
    \rho_S(0)_4&=\left(\ket{0}\bra{0}\right)^{\otimes 4},\nonumber \\
    \rho_S(0)_5&=\left(\ket{1}\bra{1}\right)^{\otimes 4}.
\end{align}%
The initial state of the bath is taken to be the same as that of the system’s state for each case. Figure~\ref{fig:LEchoU_TFIM_varyRhoS0} shows that when the initial state is maximally mixed [$\rho_S(0)_2$], the value of the Loschmidt echo remains one throughout, and only for initial pure states [all other states in Eq.~\eqref{eq_initial_states}], the Loschmidt echo exhibits sensitivity to chaos. 

For the Loschmidt echo, the general trend for a chaotic system is that it decays with irregular oscillations. For an integrable system, it shows perfect oscillations when the perturbation to the Hamiltonian is zero. The more the model enters into its chaotic regime, the more the periodic nature of the Loschmidt echo is destroyed. Figure~\ref{fig:LEchoU_TFIM_vary_pert}(a) illustrates this phenomenon. When a perturbation to the Hamiltonian is added, then even in the integrable regime, we see in Fig.~\ref{fig:LEchoU_TFIM_vary_pert}(d) decay of the Loschmidt echo.

\section{Speed of information scrambling}\label{operator growth}
Previously, we have discussed information scrambling in a spin chain system in terms of out-of-time ordered correlator ($\mathcal{F}$-OTOC). In this section, we shall talk about information scrambling in the light of commutator growth. A local perturbation applied at a site evolves to other sites. For chaotic Hamiltonians, the information spreads fast, causing a loss of memory of the perturbation. In other words, initially, two local commuting operators at different sites share no information. As the perturbation caused by the local operator grows through the spin chain, non-commutation sets in, and operators at two different sites start sharing information. A measure of information scrambling for an open system is the operator norm of the commutator $[A_S(t), B_S]$, which is defined as
\begin{align}
    O(t)=||[ A_S(t), B_S(0) ]||_{\rm op}.
\end{align}
The upper limit of this quantity is called the Lieb Robinson bound ~\cite{Hastings_LR_bound, Lieb1972, LiebRobSwingle_PhysRevLett.117.091602},
\begin{align}\label{LRbound}
    \|[A_S(x,t), B_S(0)]\|_{\rm op} \leq K_0 \|A_S\|_{\rm op} ~\|B_S\|_{\rm op} e^{-(|x|-v_{\text{LR}}t)/\xi_0},
\end{align}
where \( K_0 \) and \( \xi_0 \) are constants.~\( \|\cdot\|_{\rm op} \) indicates the operator norm which is the maximum eigenvalue of \( [A_S(x,t), B_S(0)]^{\dagger} [A_S(x,t), B_S(0)] \), and \( v_{\text{LR}} \) is the Lieb-Robinson velocity. We calculate and plot $O(t)$ for the TC and TFI models. The information spreads through a chaotic system ballistically, forming a light cone with a velocity bounded by the Lieb-Robinson velocity.
\begin{figure}
    \centering
    \includegraphics[width=1\linewidth]{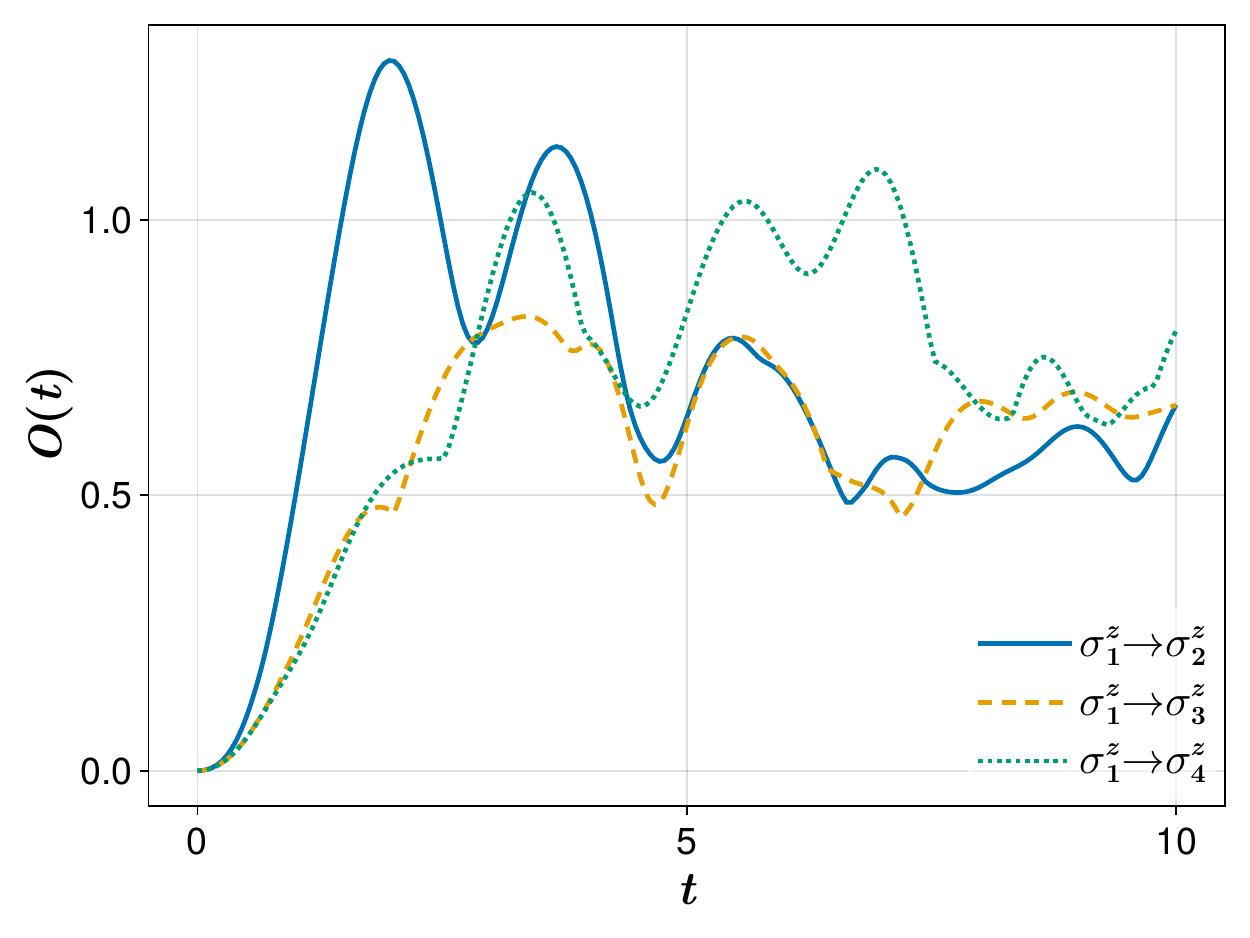}
    \caption{$O(t)$'s are plotted for TC model. The three plots are for three different growths of perturbation from spin 1 to other spins. Other parameters are $j_{TC}=\frac{\lambda}{2\sqrt{N}}=0.5$ and the interaction strength $j_s=0.5$  of the nearest neighbor interaction in the system spin chain. The other parameters are $\omega_0=2,~\omega_c=2,~T=10$.}
    \label{fig:LiebRob_TC_Szvary}
\end{figure}

In Fig.~\ref{fig:LiebRob_TC_Szvary}, we plot the variation of $O(t)$ for the TC model. Since $O(t)$ is constructed as the growth of the commutator, it starts from zero, where two distant local operators commute with each other. Initially, an exponential rise is seen, where the rise is different for three sets of initial and final spins. The bound of $O(t)$ in~\eqref{LRbound} shows that the more distant the spins are, the more suppressed the exponential rise of $O(t)$ with time would be. We observe in Fig.~\ref{fig:LiebRob_TC_Szvary} that the steepest rise is between the closest spin sites. However, the information light cone of the scrambling is not apparent in this case because of the direct decay of each of the local operators under the open system evolution. It is also seen that the $O(t)$ doesn't saturate. Due to the non-Markovian nature of the environment, it fluctuates on time scales $\simeq \mathcal{O}(N)$, $N$ being 4 in this case.

\begin{figure}
    \centering
    \includegraphics[width=1\linewidth]{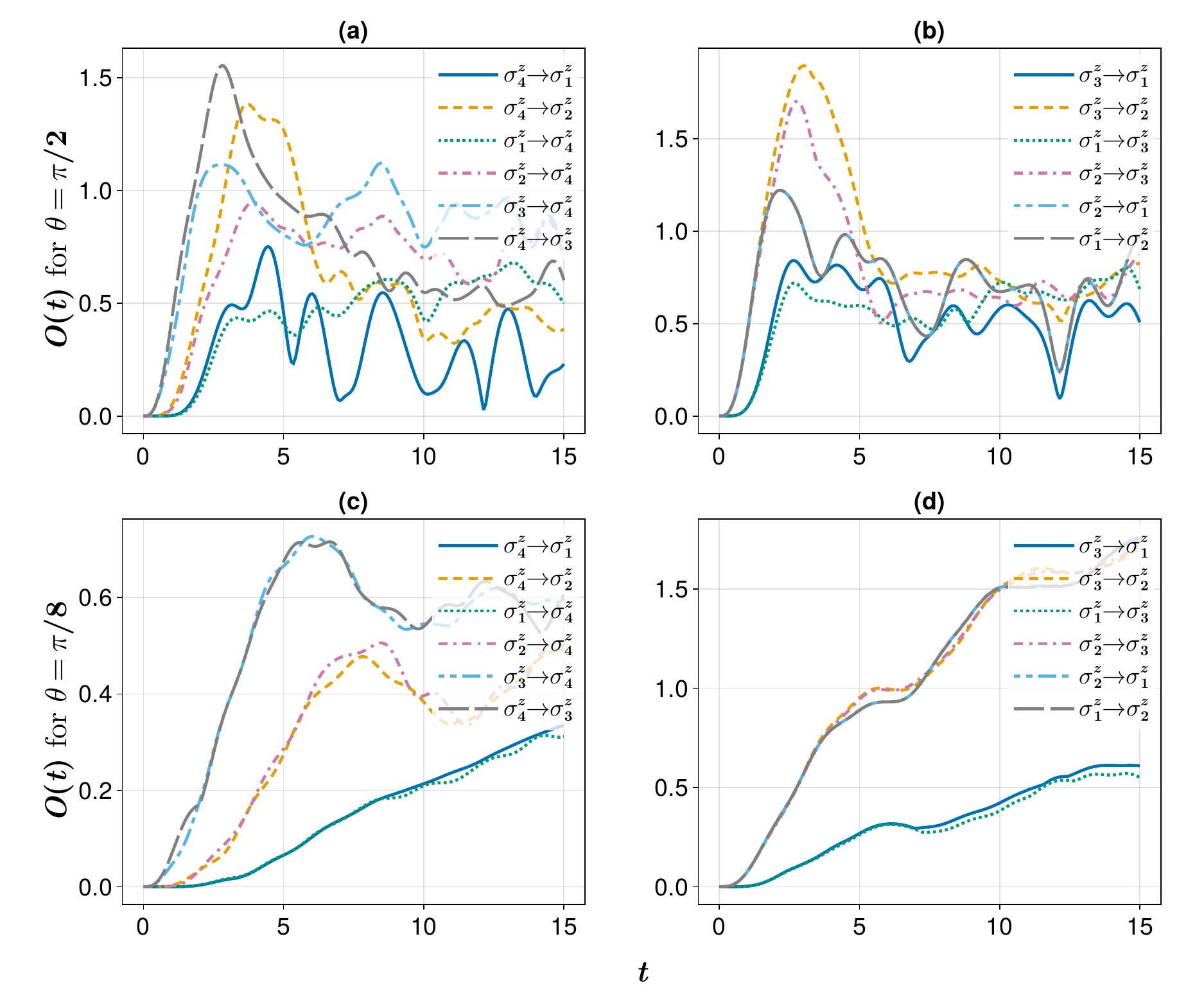}
    \caption{$O(t)$ for TFIM model of four system Ising spins interacting with a four Ising spin bath. For (a) and (b), $\theta=\pi/2$ and for (c) and (d) $\theta=\pi/8$.  The other parameters chosen are $\mathcal{B}=J=0.5$.}
    \label{fig:LiebRob_for_TFIM_pi2_8}
\end{figure}

In the TFIM model, each of the spins is not coupled to the environment. We see the variation of $O(t)$ for the TFI model at two angles $\theta=\pi/2,\pi/8$ in Fig.~\ref{fig:LiebRob_for_TFIM_pi2_8}.
For $\theta=0$, not depicted in the figure, the model is integrable, and the zero growth of the operator norm of the commutator $O(t)$ suggests the non-chaotic nature of the model at $\theta = 0$.
For the non-trivially integrable TFIM at $\theta=\pi/2$, and also for the non-integrable TFIM at $\theta=\pi/8$, we observe $O(t)$ has an almost exponential rise at small times; particularly for the non-trivially integrable case, $\theta = \pi/2$, the rise is steep. For $\theta = \pi/8$, the decay from the initial rise is not as sharp as compared to $O(t)$ at higher angles. The sharpness of the decay from the initial rise is the most prominent at $\theta=\pi/2$. For all cases at both angles, the light cones emerge perfectly, i.e., $O(t)$ remains zero longer for farther spins, making it evident that information transfer between spins at greater distances requires longer times. The initial commutativity of the local operators renders $O(t)$ constant at zero before it increases at the onset of scrambling. The duration for which $O(t)$ remains zero varies almost linearly with the distance between spins. This means the information is spreading through the system ballistically. Owing to the Lieb Robinson bound, at smaller times, we see steeper exponential growth occur for nearer spins. 

For all the above chaotic models, again, we do not see complete saturation of $O(t)$, which implies that the system doesn't fully forget the initial perturbations. The non-Markovian nature of the bath causes the revival of the memory of those initial perturbations.

The trend followed by the $\mathcal{F}-$OTOC described in the last paragraph of Sec.~\ref{nmgad} is also followed by the commutator growth $O(t)$, except for the fact that what is seen as the initial drop in one case of $\mathcal{F}-$OTOCs is replaced by an initial rise from zero. The $O(t)$ is zero for an integrable model, providing the baseline.

\section{Conclusions}\label{conclusion}
We investigated the scrambling of information in open quantum systems by studying the interferometric out-of-time-ordered correlator ($\mathcal{F}$-OTOC), Loschmidt echo, and speed of information scrambling calculated using the operator norm of the appropriate commutator. Our focus was on understanding how the non-Markovian nature of the system influences this process. To explore these effects, two models were considered: the Tavis-Cummings (TC) model and a generalized variant of the Ising model, namely the tilted field Ising model (TFIM). In the TC model, where the bath impacts all the qubits, the light cone, indicating ballistic transfer of information through the spins, was not apparent in the $\mathcal{F}$-OTOC. It reappeared in the corrected $\mathcal{F}$-OTOC, which is designed to suppress the dissipative effects.
However, in the TFI model, a light cone emerged for both $\mathcal{F}$-OTOC and the corrected $\mathcal{F}$-OTOC. Both models depicted late-time revivals in $\mathcal{F}$-OTOC, indicative of the non-Markovian nature of the evolution. The study suggested that integrability and quantum chaos have an intricate relationship. The TFIM at $\theta = \pi/2$, though integrable, still exhibited chaotic behavior, and the intermediary angles, even though non-integrable, could show non-chaotic behavior. The counterintuitive chaotic nature of TFIM at $\theta = \pi/2$ is interesting as it is non-trivially integrable and demands a deeper exploration, which we hope to study in the future.
Additionally, we analyzed the Loschmidt echo, which characterizes irreversible effects due to perturbations in the system Hamiltonian and shares a structural similarity with the $\mathcal{F}$-OTOC, making it relevant in the present context. The Loschmidt echo suffers decay in the TC model primarily due to dissipation, and in the TFIM, except for $\theta = 0$, due to the chaotic nature of the dynamics. Interestingly, for $\theta=0$, the Loschmidt echo decays if a perturbation that is non-commuting with the system Hamiltonian is added to the backward evolution.
In systems with local interactions, such as spin models, the spread of information is bounded by the Lieb-Robinson (LR) bound, which provides an upper limit on the operator norm of the commutator growth. Here, we observed this for the TC and the TFI models. The commutator growth showed an approximate exponential rise for both models. Particularly, in the TFI model, the light cone emerged in the operator growth. Operator spreading between two given spin sites showed distinct profiles owing to their positions in the spin chain. Further, the operator growth didn't saturate and showed revivals because of the non-Markovian nature of the evolution.  
In summary, our study sheds light on the interplay between information scrambling, irreversible effects, quantum chaos, and non-Markovianity in open quantum systems.

\appendix
\section{Derivation of $\mathcal{F}$-OTOC for open system}
The basic interferometric scheme (illustrated in \cite{SwingleInfoScram}) of calculating $\mathcal{F}(t)$ can be formulated in the following way.
\begin{enumerate}
    \item Two interferometric arms are taken as $\ket{\psi}_S\ket{1}_c$ and $\ket{\psi}_S\ket{0}_c$.

    \item $B_S$ acts on $\ket{\psi}_S\ket{1}_c$ at time $t=0$ and then $A_S$ acts on it at a later time $t$, rendering $A_S(t)B_S\ket{\psi}_S\ket{1}_c$.

    \item $A_S$ acts on $\ket{\psi}_S\ket{0}_c$ at time $t$ and then $B_S$ acts on it at an earlier time $t=0$, rendering $B_SA_S(t)\ket{\psi}_S\ket{0}_c$.

    \item The two states containing different orders of operation of the local operators are then made to interfere with each other, $\frac{1}{\sqrt{2}}((B_SA_S(t)\ket{\psi}_S)\ket{0}_c+(A_S(t)B_S\ket{\psi}_S)\ket{1}_c)$.

    \item The control qubit is measured in the $x$ and $y$ bases to get the real and imaginary parts of the OTOC. The real part of the OTOC is $\mathcal{F}(t)$.
\end{enumerate}
The final state after the interference is shown below to illustrate the reason for measuring it in the $x$ basis in order to obtain the $\mathcal{F}(t)$.
\begin{align}\label{rhof}
    \rho_f=\frac{1}{2}\begin{pmatrix}
        B_SA_S(t)\rho_S(0)A^{\dagger}_S(t)B^{\dagger}_S & B_SA_S(t)\rho_S(0)B^{\dagger}_S(t)A^{\dagger}_S \\
        A_SB_S(t)\rho_S(0)A^{\dagger}_S(t)B^{\dagger}_S & A_SB_S(t)\rho_S(0)B^{\dagger}_S(t)A^{\dagger}_S 
    \end{pmatrix}.
\end{align}
In \eqref{rhof}, the diagonal terms are time-ordered, and the non-diagonal terms are out-of-time ordered. To sort out the non-diagonal terms, $\sigma_c^x=(\ket{0}\bra{1}_c+\ket{1}\bra{0}_c)$ is measured. Exploiting the fact that the non-diagonal terms are the complex conjugate of each other, we finally obtain the $\mathcal{F}(t)$ as the real part of one of the out-of-time ordered terms.

Now, let us calculate $\mathcal{F}(t)$ in mathematical detail. The initial state $\rho_S(0)$ is associated with a $\ket{+}_c$ state of the control qubit, and then the scheme in~\eqref{scheme} is applied. The initial state is $\rho_1=\rho_S(0)\otimes \frac{1}{2}\left(|0\rangle\langle 0|_c+|0\rangle\langle 1|_c+|1\rangle\langle 0|_c+|1\rangle\langle 1|_c \right)$
The first step of Eq.~\eqref{scheme}, $\mathcal{S}_1=\mathcal{C}(\mathbf{I}_S\otimes |0\rangle\langle 0|_c + B_S \otimes |1\rangle\langle 1|_c ) $ is applied and we obtain,
\begin{align}
    \rho_2&=\mathcal{S}_1\cdot\rho_1 \nonumber \\
          &=\frac{1}{2} \Bigg\{ \rho_S(0) \otimes |0\rangle\langle 0|_c + \rho_S(0) B_S^{\dagger} \otimes |0\rangle\langle 1|_c \\
&+ B_S \rho_S(0) \otimes |1\rangle\langle 0|_c + B_S \rho_S(0) B_S^{\dagger} \otimes |1\rangle\langle 1|_c \Bigg\}.
\end{align}
After the second operation, the state becomes,
\begin{align}
    \rho_3&=\mathcal{S}_2\cdot\rho_2 \nonumber \\
          &=\frac{1}{2} \Bigg\{ \xi_f(t) \rho_S(0) \otimes |0\rangle\langle 0|_c + \xi_f(t)\rho_S(0) B_S^{\dagger} \otimes |0\rangle\langle 1|_c \nonumber \\
&+ \xi_f(t) B_S \rho_S(0) \otimes |1\rangle\langle 0|_c + \xi_f(t) B_S \rho_S(0) B_S^{\dagger} \otimes |1\rangle\langle 1|_c \Bigg\}.
\end{align}
Now, we illustrate the action of the superoperators, which are the likes of $\mathcal{S}_2=\xi_f(t)\otimes\mathcal{I}_c$, where $\mathcal{I}_c$ is not an identity operator but an identity superoperator. 
The effect of $\xi_f(t)$ on the state $\rho_S(0)\otimes |0\rangle\langle0|_c$ is
\begin{align}
    &(\xi_f(t)\otimes\mathcal{I}_c)\cdot(\rho_S(0)\otimes |0\rangle\langle0|_c) \nonumber \\
    &={\rm Tr_E}(e^{-i(H_f\otimes I_c)t} (\rho_S(0) \otimes \rho_E(0) \otimes |0\rangle\langle0|_c ) e^{i(H_f\otimes I_c)t}) \nonumber \\
    &={\rm Tr_E}(e^{-iH_f t} (\rho_S(0) \otimes \rho_E(0)  ) e^{iH_f t}) \otimes |0\rangle\langle0|_c \nonumber \\
    &=\xi_f(t)\cdot\rho_S(0)\otimes  |0\rangle\langle0|_c.
\end{align}
On applying $\mathcal{S}_3=\mathcal{C}(A_S \otimes I_c)$ to $\rho_3$, we obtain
\begin{align}
    \rho_4&=\mathcal{S}_3\cdot\rho_3 \nonumber \\
          &=\frac{1}{2} \Bigg\{ A_S\xi_f(t) \rho_S(0)A_S^{\dagger} \otimes |0\rangle\langle 0|_c \nonumber \\
          &+ A_S\xi_f(t)\rho_S(0) B_S^{\dagger}A_S^{\dagger} \otimes |0\rangle\langle 1|_c \nonumber \\
&+ A_S\xi_f(t) B_S \rho_S(0)A_S^{\dagger} \otimes |1\rangle\langle 0|_c \nonumber \\
          & + A_S\xi_f(t) B_S \rho_S(0) B_S^{\dagger}A_S^{\dagger} \otimes |1\rangle\langle 1|_c \Bigg\}.
\end{align}
In the next step, $\mathcal{S}_4=\xi_b(t)\otimes\mathcal{I}_c$ is applied on $\rho_4$ as
\begin{align}
    \rho_5&=\mathcal{S}_4\cdot\rho_4 \nonumber \\
          &=\frac{1}{2} \Bigg\{ \xi_b(t) A_S\xi_f(t) \rho_S(0)A_S^{\dagger} \otimes |0\rangle\langle 0|_c \nonumber \\
          &+ \xi_b(t)A_S\xi_f(t)\rho_S(0) B_S^{\dagger}A_S^{\dagger} \otimes |0\rangle\langle 1|_c \nonumber \\
&+\xi_b(t) A_S\xi_f(t) B_S \rho_S(0)A_S^{\dagger} \otimes |1\rangle\langle 0|_c \nonumber \\
          & + \xi_b(t)A_S\xi_f(t) B_S \rho_S(0) B_S^{\dagger}A_S^{\dagger} \otimes |1\rangle\langle 1|_c \Bigg\},
\end{align}
and $\rho_5$ is obtained.
Here, $\xi_b(t)$ acts similarly as $\xi_f(t)$ as illustrated above. Now, the final operation $\mathcal{S}_5=\mathcal{C}(B_S\otimes |0\rangle\langle 0|_c + \mathbf{I}_S \otimes |1\rangle\langle 1|_c )$ renders $\rho_f$,
\begin{align}
    \rho_f&=\mathcal{S}_5\cdot\rho_5 \nonumber \\
          &=\frac{1}{2} \Bigg\{B_S \xi_b(t) A_S\xi_f(t) \rho_S(0)A_S^{\dagger} B_S^{\dagger}\otimes |0\rangle\langle 0|_c \nonumber \\
          &+ B_S\xi_b(t)A_S\xi_f(t)\rho_S(0) B_S^{\dagger}A_S^{\dagger}   \otimes |0\rangle\langle 1|_c \nonumber \\
&+\xi_b(t) A_S\xi_f(t) B_S \rho_S(0)A_S^{\dagger}B_S^{\dagger} \otimes |1\rangle\langle 0|_c \nonumber \\
          & + \xi_b(t)A_S\xi_f(t) B_S \rho_S(0) B_S^{\dagger}A_S^{\dagger} \otimes |1\rangle\langle 1|_c \Bigg\}.
\end{align}
We see that the time-ordered terms that are represented by the diagonal elements of the control qubit cancel out when measured in the $x$ basis of the control qubit, and we are left with two out-of-time-ordered terms.

Finally, to obtain the $\mathcal{F}$-OTOC, $\mathcal{F}(t, A, B)$, tracing with respect to $\mathbf{I}_S\otimes \sigma_c^x$ is done. Two terms that survive this operation are
\begin{align}\label{FOTOC_twoterm}
    \mathcal{F}(t,A,B)&={\rm Tr}\left\{ \left(\mathbf{I}_S\otimes ( |0\rangle\langle1| + |1\rangle\langle 0|) \right)\cdot \rho_f \right\} \nonumber \\
   &= {\rm Tr} \left[ \xi_b(t) A_S\xi_f(t) B_S \rho_S(0)A_S^{\dagger}B_S^{\dagger}\right. \nonumber \\
   &+\left. B_S\xi_b(t)A_S\xi_f(t)\rho_S(0) B_S^{\dagger}A_S^{\dagger}\right].
\end{align}
Let us analyze the first term,
\begin{align} \label{1st_term}
   &{\rm Tr} \left( \xi_b(t) A_S\xi_f(t) B_S \rho_S(0)A_S^{\dagger}B_S^{\dagger} \right)\nonumber \\
   &={\rm Tr} \left( B_S^{\dagger} \xi_b(t) A_S\xi_f(t) B_S \rho_S(0)A_S^{\dagger} \right),
\end{align}
which, using the cyclic property of trace, can be written as 
\begin{align}
    &{\rm Tr} \left( \xi_b(t) A_S\xi_f(t) B_S \rho_S(0)A_S^{\dagger}B_S^{\dagger} \right) \nonumber \\ 
    &= {\rm Tr} \left( B_S^{\dagger} \xi_b(t) \cdot \Big( A_S\xi_f(t) B_S \rho_S(0)A_S^{\dagger} \Big) \right),
\end{align}
whereupon shifting to the adjoint map $\xi_b^{\dagger}(t)$, which acts on the operator $B_S^{\dagger}$ and not the bracketed density matrix next to it, we can write
\begin{align}
 &{\rm Tr} \left( \xi_b(t) A_S\xi_f(t) B_S \rho_S(0)A_S^{\dagger}B_S^{\dagger} \right) \nonumber \\
    &={\rm Tr} \left( \xi^{\dagger}_b(t) B_S^{\dagger}  \cdot \Big( A_S\xi_f(t) B_S \rho_S(0)A_S^{\dagger} \Big) \right)
\end{align}
We now take the second term in~\eqref{FOTOC_twoterm} and apply complex conjugation to the term inside the trace operation,
\begin{align}\label{A11}
     &\left(B_S\xi_b(t) \left( A_S\xi_f(t) \left( \rho_S(0) B_S^{\dagger} \right) A_S^{\dagger} \right) \right)^{\dagger} \nonumber \\
     &=\left( \xi^{\dagger}_b(t) B_S \cdot \left( A_S\xi_f(t) \left( \rho_S(0) B_S^{\dagger} \right) A_S^{\dagger} \right) \right)^{\dagger} \nonumber \\
     &=\left( \xi^{\dagger}_b(t) B_S^{\dagger} \cdot \left( A_S\xi_f(t) \left(B_S \rho_S(0)  \right) A_S^{\dagger} \right) \right),
\end{align}
we arrive at the first term. It is evident that the second term is the complex conjugate of the first term. We finally obtain $\mathcal{F}(t, A, B)$ as the real part of the trace of the first term
\begin{align}
    \mathcal{F}(t,A,B)=\Re \Big[ {\rm Tr} \left( \xi^{\dagger}_b(t) B_S^{\dagger} \cdot \left( A_S \left( \xi_f(t) \left(B_S \rho_S(0)  \right) \right) A_S^{\dagger} \right) \right) \Big].
\end{align}

\section{Non-Markovianity measure}
Here, we discuss the Breuer-Laine-Piilo (BLP) measure of non-Markovianity~\cite{blp_measure} for the TC and TFI models discussed in the text. In this measure, we calculate the trace distance 
\begin{align}
    T(\rho_1, \rho_2) = \frac{1}{2}{\rm Tr}\sqrt{(\rho_1 - \rho_2)^\dagger(\rho_1 - \rho_2)},
\end{align}
between two states $\rho_1$ and $\rho_2$ of the system at any time $t$. A revival in the variation of this trace distance is an indicator of the non-Markovian evolution of the system.  
\begin{figure}[h]
    \centering
    \includegraphics[width=1\linewidth]{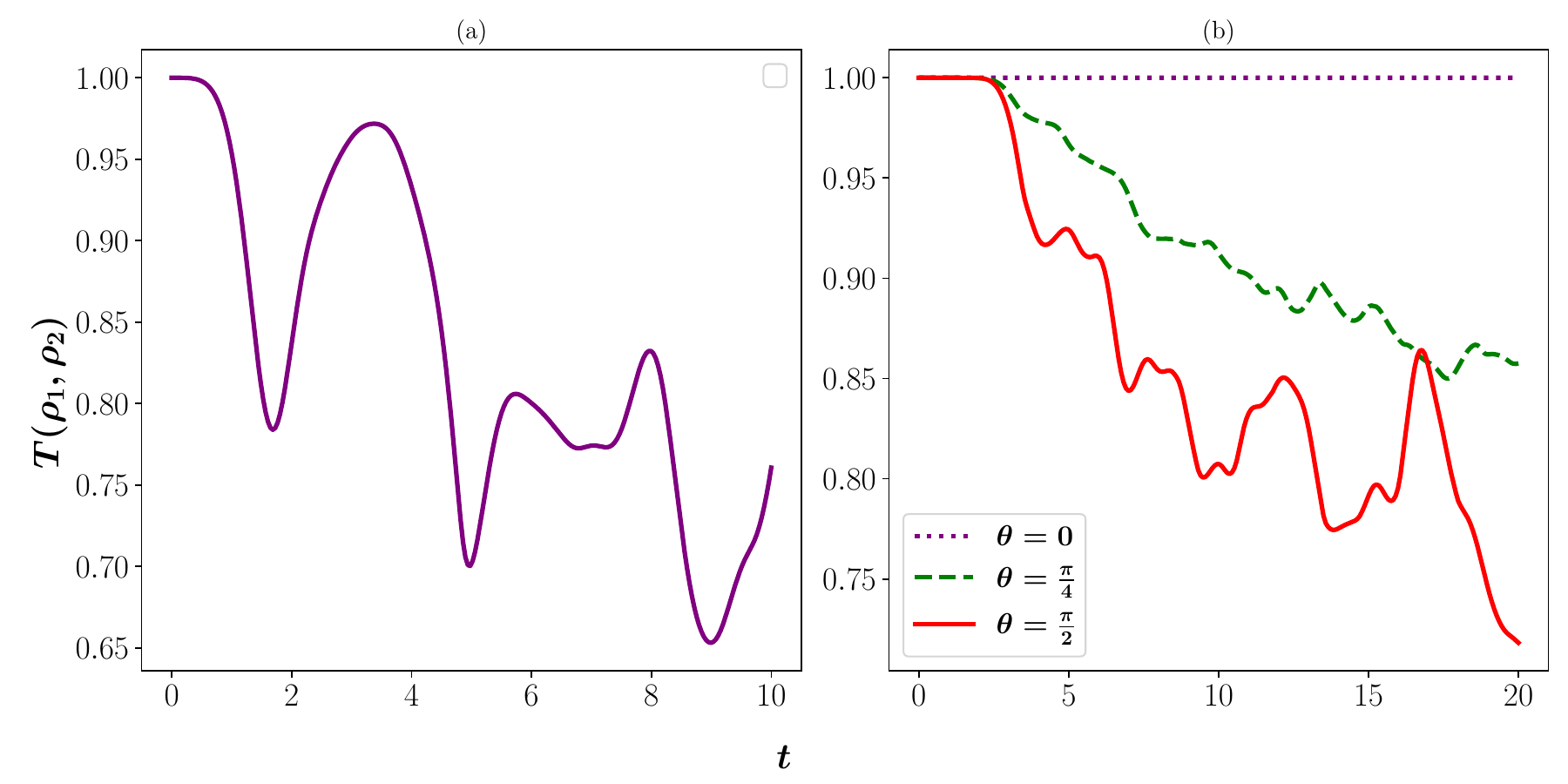}
    \caption{Variation of the trace distance $T(\rho_1, \rho_2)$ with time for the (a) TC model and (b) TFI model. In (a), $\omega_0 = 2, \omega_c = 2.5, j_s = 0.5, \lambda = 1.5$, and $T = 1$. In (b), $\mathcal{B} = 0.5$ and $J = 0.75$. }
    \label{fig_non_Markovian_plot}
\end{figure}
The variation of trace distance $T(\rho_1, \rho_2)$ for the TC and the TFI models is plotted in Fig.~\ref{fig_non_Markovian_plot}. In both cases, we take the initial states of the system to be $\rho_1 = \ket{+}^{\otimes 4}$ and $\rho_2 = \ket{-}^{\otimes 4}$. It can be observed that in the TC model, the trace distance has revivals in its evolution, indicating the non-Markovian nature of the system. For the TFIM, at $\theta=0$, the trace distance is constant at one. However, TFIM, at other angles, shows revivals in trace distance, indicative of non-Markovian evolution.  

\bibliographystyle{apsrev4-1}
\bibliography{reference}  

\end{document}